\documentclass[8.5pt,twoside,twocolumn]{article}
\pdfoutput=1
\usepackage[super,sort&compress,comma]{natbib} 
\usepackage{mhchem}
\usepackage{times,mathptmx}
\usepackage{sectsty}
\usepackage{balance} 

\usepackage{graphicx} 
\usepackage{lastpage}
\usepackage[format=plain,justification=raggedright,singlelinecheck=false,font=small,labelfont=bf,labelsep=space]{caption} 
\usepackage{fancyhdr}
\begin{document}

\thispagestyle{plain}
\fancypagestyle{plain}{
\fancyhead[L]{\includegraphics[height=8pt]{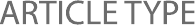}}
\fancyhead[C]{\hspace{-1cm}\includegraphics[height=20pt]{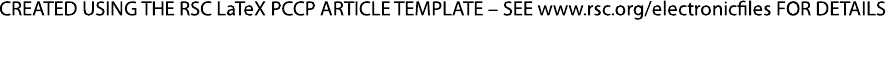}}
\fancyhead[R]{\includegraphics[height=10pt]{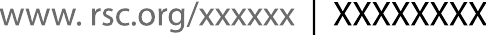}\vspace{-0.2cm}}
\renewcommand{\headrulewidth}{1pt}}
\renewcommand{\thefootnote}{\fnsymbol{footnote}}
\renewcommand\footnoterule{\vspace*{1pt}%
\hrule width 3.4in height 0.4pt \vspace*{5pt}} 
\setcounter{secnumdepth}{5}

\makeatletter 
\def\subsubsection{\@startsection{subsubsection}{3}{10pt}{-1.25ex plus -1ex minus -.1ex}{0ex plus 0ex}{\normalsize\bf}} 
\def\paragraph{\@startsection{paragraph}{4}{10pt}{-1.25ex plus -1ex minus -.1ex}{0ex plus 0ex}{\normalsize\textit}} 
\renewcommand\@biblabel[1]{#1}            
\renewcommand\@makefntext[1]%
{\noindent\makebox[0pt][r]{\@thefnmark\,}#1}
\makeatother 
\renewcommand{\figurename}{\small{Fig.}~}
\sectionfont{\large}
\subsectionfont{\normalsize} 

\fancyfoot{}
\fancyfoot[LO,RE]{\vspace{-7pt}\includegraphics[height=9pt]{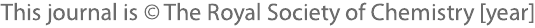}}
\fancyfoot[CO]{\vspace{-7.2pt}\hspace{12.2cm}\includegraphics{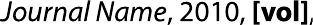}}
\fancyfoot[CE]{\vspace{-7.5pt}\hspace{-13.5cm}\includegraphics{RF}}
\fancyfoot[RO]{\footnotesize{\sffamily{1--\pageref{LastPage} ~\textbar  \hspace{2pt}\thepage}}}
\fancyfoot[LE]{\footnotesize{\sffamily{\thepage~\textbar\hspace{3.45cm} 1--\pageref{LastPage}}}}
\fancyhead{}
\renewcommand{\headrulewidth}{1pt} 
\renewcommand{\footrulewidth}{1pt}
\setlength{\arrayrulewidth}{1pt}
\setlength{\columnsep}{6.5mm}
\setlength\bibsep{1pt}

\twocolumn[
  \begin{@twocolumnfalse}
\noindent\LARGE{\textbf{Coarse-graining DNA for simulations of DNA nanotechnology}}
\vspace{0.6cm}

\noindent\large{\textbf{
Jonathan P.\ K.\ Doye,$^{\ast}$\textit{$^{a}$} 
Thomas E.\ Ouldridge,\textit{$^{b}$} 
Ard A.\ Louis,\textit{$^{b}$} 
Flavio Romano,\textit{$^{a}$} 
Petr \v{S}ulc,\textit{$^{b}$}
Christian Matek,\textit{$^{b}$} 
Benedict E.\ K.\ Snodin,\textit{$^{a}$}
Lorenzo Rovigatti,\textit{$^{c}$}
John S.\ Schreck,\textit{$^{a}$} 
Ryan M.\ Harrison\textit{$^{a}$} and  
William P.\ J.\ Smith\textit{$^{a}$}
}}
\vspace{0.5cm}

\noindent\textit{\small{\textbf{Received Xth XXXXXXXXXX 20XX, Accepted Xth XXXXXXXXX 20XX\newline
First published on the web Xth XXXXXXXXXX 200X}}}

\noindent \textbf{\small{DOI: 10.1039/b000000x}}
\vspace{0.6cm}

\noindent \normalsize{To simulate long time and length scale processes
involving DNA it is necessary to use a coarse-grained description. Here
we provide an overview of different approaches to such coarse graining,
focussing on those at the nucleotide level that allow the
self-assembly processes associated with DNA nanotechnology to be studied. 
OxDNA, our recently-developed coarse-grained DNA model, is particularly suited
to this task, and has opened up this field to systematic study by simulations.
We illustrate some of the range of DNA nanotechnology systems to which the
model is being applied, as well as the insights it can provide into fundamental
biophysical properties of DNA.
} \vspace{0.5cm} \end{@twocolumnfalse} ]

\section{Introduction}
\label{sect:intro}

\footnotetext{\textit{$^{a}$Physical \& Theoretical Chemistry Laboratory, Department of Chemistry, University of Oxford, South Parks Road, Oxford, OX1 3QZ, UK. }}
\footnotetext{\textit{$^{b}$Rudolf Peierls Centre for Theoretical Physics, 1 Keble Road, Oxford, OX1 3NP, UK. 
}}
\footnotetext{\textit{$^{c}$ Dipartimento di Fisica, Universit\`{a} di Roma La Sapienza, Piazzale A. Moro 2, 00185 Roma, Italy.
}}

DNA is one of the fundamental molecules of biology, enjoying iconic status in
its role as the information storage medium for the genome.  More recently, DNA
has also become a startlingly successful nanoengineering material, with the
exponents of DNA nanotechnology having developed an impressive array of DNA
nanostructures\cite{Linko13} and nanodevices.\cite{Bath07} For both these uses
of DNA, its physico-chemical properties are vital to its role.  

For example, the intra- and intermolecular
interactions that drive base pairing and base stacking are responsible for
DNA's double helical structure. Furthermore, the selectivity of Watson-Crick
base-paring and the efficiency with which double helices can self-assemble
underlie both fundamental genomic processes, such as replication and
transcription, and the programmability of DNA that has been exploited so
successfully by DNA nanotechnology. 
On longer length scales, the mechanical properties of the double helix are
important both to how DNA can be manipulated in the cellular environment, such
as through the packaging of DNA by complexation with proteins and the control
of its global topology by enzymes that modify its local twist, and in imparting
rigidity to DNA nanostructures.

Experimentally, there is an increasing wealth of information on DNA's
biophysical properties, particularly more recently from single-molecule
experiments involving active manipulation by optical and magnetic
tweezers\cite{Bryant12,Forth13} or by passive fluorescent probing of single
molecules as they undergo dynamic processes.\cite{Zhang09b,Vafabakhsh12b}
However, the detailed microscopic mechanisms underlying the observed behaviour
in these experiments are not always clear. 

DNA nanotechnology is also a relatively young field.  It originates from the
1980s when Nadrian Seeman first proposed making artificial objects out of DNA
that could form by self-assembly.\cite{Seeman82} In the last decade this field
has seen very rapid growth, both in terms of the size of the literature, and
the complexity of the objects that have been produced.  The structures
typically consist of
double-helices joined together by single-stranded sections or by junctions at
which strands cross over between helices, but the technology has now developed
to such an extent that objects of seemingly any shape can be
assembled.\cite{Linko13} 
One of the distinctives of these DNA nanostructures is that they can be
addressed with nanometer precision, and so one recent theme has been the
decoration of such structures with, for example, nanoparticles\cite{Kuzyk12}
and other biomolecules\cite{Langecker12} in order to create functionality or
to use as substrates for single-molecule experiments.  A second attractive
feature is that structures can be made that are able to respond, usually to
the presence of other nucleic acid strands but also to other
molecules,\cite{Douglas12} to pH\cite{Modi09} and to light.\cite{You12} Such
devices have included nanotweezers,\cite{Yurke00} openable
boxes,\cite{Andersen08} walkers that can move along tracks\cite{Wickham11} and
responsive drug delivery capsules.\cite{Douglas12}

The most basic design principle for DNA nanostructures is simple; namely, that
the target structure should be the global free-energy minimum, and this can
usually be achieved if the target has the most base pairs.  However, how such
structures, particularly the larger and more complex ones, which can be made
up of tens of thousands of nucleotides and hundreds of different strands, are
able to self-assemble is still somewhat of a mystery. Moreover, there are
relatively few guiding principles for how to prevent the systems from getting
stuck in kinetic traps, and hence to improve yields. For DNA nanodevices,
well-established nearest-neighbour thermodynamic models of duplex and hairpin
stability\cite{SantaLucia04} can help provide insight into their action.
However, the rates of the different processes involved are also important.
Furthermore, these devices can involve non-trivial multi-strand complexes with
non-trivial internal loops or pseudoknots that can be subject to internal or
externally applied stresses and whose stabilities are not yet captured by
these thermodynamic models.

The above issues suggest that there could be a significant potential role for
molecular simulations in DNA nanotechnology: firstly, in providing an improved
microscopic understanding of basic DNA biophysics relevant to DNA
nanotechnology; secondly, in visualizing the mechanisms of self-assembly for
DNA nanostructures and of the action of DNA nanodevices,  and thirdly in
aiding the rational design and optimization of DNA nanotechnology systems
using the potential quantitative insights into their thermodynamics and
dynamics. So far, however, the contribution of molecular simulations to DNA
nanotechnology has been relatively minor.  

One significant obstacle is that all-atom simulations are not yet capable of
probing the large system sizes and long time scales that are typically
relevant to DNA nanotechnology systems.  Therefore, to progress the use of
simulations in this field, there is a need for ``coarse-grained'' models
that provide a simpler representation of DNA, but one which hopefully retains
enough of the essential physics to allow realistic modelling of DNA systems on
long time and length scales. However, until relatively recently, there were
few such coarse-grained models,  and coarse-graining has been more explored
for other biomolecules\cite{Takada12} such as proteins\cite{Saunders12} and
lipids.\cite{Shinoda12}
This is perhaps surprising given that, in some ways, DNA is an easier target
for coarse-graining, certainly when compared to proteins, because of the
relatively limited number of ways the bases interact --- the physical
behaviour of DNA is dominated by the stacking of the bases and the pairing of
complementary bases through hydrogen-bonding.  Only in the last few years have
the number of coarse-grained DNA models increased
significantly,\cite{dePablo11,Lankas11,Potoyan13}  but only a few of these
have the potential to realistically model self-assembly processes associated
with DNA nanotechnology systems. Notably, the first published example of such
an application was only in 2010 when nanotweezers similar to those developed
by Yurke {\it et al.}\cite{Yurke00} were simulated.\cite{Ouldridge10}

Here we provide a selective overview of coarse-grained models of DNA in Section
\ref{sect:CG} focussing down on those that have the requisite properties to
allow the simulation of DNA nanotechnology and highlighting the oxDNA model
that we have recently developed for this purpose in Section \ref{sect:oxDNA}.
In Section \ref{sect:sims} we discuss some of the technical issues that are
relevant when computing the thermodynamic and dynamic properties of such
models by molecular simulation, before illustrating the power and range of
applicability of oxDNA by showing a number of nanotechnological and 
biophysical systems to which the model has been applied in Section
\ref{sect:results}.

\section{Coarse-grained DNA models: a selective overview}
\label{sect:CG}

When seeking to understand a particular property of DNA by theory or
simulation, one of course wants to choose a level of description that captures
enough of the physics to allow the question of interest to be addressed,
whilst also remaining sufficiently simple that such a calculation is tractable.  For
this reason, there is a whole spectrum of approaches to describe DNA involving
differing levels of detail. 

At one end of this spectrum are polymer theories such as the worm-like chain
model, which describes DNA as a polymer with a certain bending modulus (and
sometimes finite twist and extensional moduli).  This model is particularly
suited to looking at the elastic properties of DNA on long length scales. For
example, it provides a good description of the response of double-stranded DNA
to low forces, capturing its entropic elasticity.\cite{Bustamante94} However,
such a description begins to break down for mechanical properties that involve
more extreme perturbations in the structure. For example, the worm-like chain
model can describe the ``$j$-factors'' for the cyclization of long DNA
duplexes, but fails for short duplexes where cyclization is most likely
achieved by local buckling rather than homogeneous bending.\cite{Peters10}

At the other end of the spectrum, certain properties can probably only be
understood using models that have an all-atom representation of DNA, because
the property reflects the detailed local geometry of the molecule. Although the
achievements of all-atom simulations of DNA are becoming increasingly
impressive,\cite{Laughton11,Perez12} due both to advances in the force-fields
used and to greater computer power, such methods are inevitably limited to
either relatively short time scales or relatively small system sizes. 

In between these two extremes are a whole range of different coarse-grained
models of varying complexity. 
As we wish to consider models that would allow 
DNA nanotechnology systems to be studied, 
in particular the self-assembly, structure and mechanical properties of DNA
nanostructures, and the action of DNA nanodevices, we should first consider the
requirements this imposes on such a model. The first and most basic requirement
is that the model has a realistic description of the three-dimensional geometry
of DNA. 
Secondly, the self-assembly of virtually all such DNA nanostructures and the
action of most of the DNA nanodevices is driven by hybridization, and so it is
important that this transition is well described. On a thermodynamic level this
includes being able to reproduce the melting points and transition widths of
duplex DNA.
Thirdly, a good description of the mechanical properties of both
single-stranded DNA (ssDNA) and double-stranded DNA (dsDNA) is required if the
structural and mechanical properties of DNA nanostructures are to be
reproduced, in particular, the rigidity of the duplex and the relative
flexibility of ssDNA.  The latter is particularly important, as virtually all
DNA nanostructures make use of the ability of the DNA backbone to bend back on
itself at junctions connecting different double helical sections of the
structures.
It can also be important to have a good description of the extensibility of
ssDNA, since sometimes the operation of nanodevices involves externally applied
or internally generated tension.\cite{Ouldridge13,Sulc13} 
Fourthly, it must be feasible to simulate the long time scales associated with
diffusion of strands in solution and with significant conformational
rearrangements. 

These consideration rule out many possible models. For example, all-atom
models are simply too computationally expensive to allow the time scales
associated with self-assembly to be simulated. The same is probably true of
those coarse-grained models that retain an explicit description of the
solvent.\cite{DeMille11} Similarly, the need to describe hybridization rules
out those coarse-grained models which are intended just to describe dsDNA and
do not
dissociate.\cite{Mergell03,Tepper05,Becker07,Becker09,Savelyev09,Savelyev10,Gonzalez13} 
Therefore, the minimal unit must at most be at the level of a single
nucleotide.  It is interesting to note that Lanka\v{s} {\it et al} found that the
level of the base also provides a more appropriate level to capture the local
dynamics of DNA than that of the base pair.\cite{Lankas09}

Models that have too reduced a representation of DNA geometry will also be
unsuitable. For example, the Peyrard-Bishop family of models have been
extensively used to study melting processes in DNA, their simplicity allowing
particularly long length scales to be explored,\cite{Peyrard04} but their
lack of three-dimensional structure makes them inappropriate for our task.
Similarly, although ``ladder models'' in which dsDNA is not helical,
\cite{SalesPardo05,Starr06,Araque11,Svaneborg12,Svaneborg12b} might be able to
give qualitative insights into the nature of self-assembly for DNA-like
polymers that have complementary base-pairing
interactions,\cite{Ouldridge09,Svaneborg12c}
they are not suitable for detailed quantitative analysis. For example, the
mechanical properties of dsDNA in the model are too anisotropic; in
particular, fluctuations out of the plane of the ``ladder'' are too facile.
However, their simplicity allows particular large systems, such as DNA-coated
nanoparticles, to be studied.\cite{Dei10}

Thus, these exclusions leave coarse-grained DNA models that can undergo melting
and have a reasonable description of basic DNA geometry. Although the
development and use of coarse-grained models of DNA has generally lagged behind
that for other biomolecules, such as proteins or lipids, there are a rapidly
increasing number of models at this
level.\cite{Drukker00,Drukker01,Mielke05,Knotts07,Trovato08,Sambriski09,Sambriski09c,Kenward09,MorrisAndrews10,Doi10,Dans10,Tito10,Allen11,Freeman11,Linak11,Kikot11,Savin11,Hsu12,Ouldridge10,Ouldridge11,Edens12,Maciejczyk10,He13,Cragnoli13}
The distinctions between these models that we wish to highlight are the
``philosophies'' behind the models and the ways that the component interactions
are represented.  

The first major question is to which data should the model be fitted, and there
are two main approaches.  The first, which we term the ``bottom-up'' approach,
is to fit the model to results (typically correlation functions) from a
finer-grained model, which is usually an empirical all-atom model (e.g. AMBER,
CHARMM),\cite{Savelyev09,Savelyev10,Kikot11,Savin11,Gonzalez13,Edens12,Maciejczyk10,He13}
although fitting to energies obtained from electronic structure calculations
(albeit in the absence of water)\cite{Hsu12} has also been performed. The
second ``top-down'' approach is to directly fit the parameters of a physically
motivated potential to reproduce measured experimental properties, e.g.\
melting transitions, persistence lengths, and elastic moduli. 

Although the bottom-up approach may provide a more direct connection to models
with finer detail, there are a number of potential issues to bear in mind with
the bottom-up approach. Firstly, it is dependent on the quality of the
fine-grained description, but for many properties in which one might be
interested, e.g.\ DNA melting points, it is often unknown how well the
fine-grained description reproduces these properties, simply because they are
currently too computationally expensive to calculate at that level. Secondly,
although some methods of coarse-graining may have more rigour than others,
there is no unique approach to achieve the coarse-graining, and
``representability'' problems are a ubiquitous feature of such
coarse-graining.\cite{Louis02} 

The second major question is how to represent the interactions. It is
probably most common to use forms that aim to capture some of the basic
structural features of the interactions. Therefore, often the potential is
broken up into terms that represent base stacking, the hydrogen-bonding between
complementary base pairs, a backbone potential, excluded volume, electrostatic
interactions between charged groups, and perhaps cross-stacking (interactions
between diagonally opposite bases in the duplex), coaxial stacking (stacking
interactions between bases that are adjacent in a helix, but are not neighbours
along a chain) and a solvation term. 

There are number of things to be born in mind when considering the interactions
in a coarse-grained model. Firstly, the interaction terms are in principle free
energies, not just energies.\cite{Everaers07,Louis02} By definition,
coarse-graining, by reducing the number of degrees of freedom, reduces the
total entropy of the system. However, this is not a problem in itself, because
we are usually concerned with entropy differences, e.g.\ between stacked and
unstacked bases, rather than absolute entropies. Nevertheless, there is no
guarantee that these relative entropy differences will be preserved by the
coarse-graining, and so consideration should be given to the potential entropic
components, and hence temperature dependence, of the effective interactions.
Most coarse-grained potentials ignore this issue and are 
temperature independent.

For some degrees of freedom that are coarse-grained away, there
is likely to be little effect on the relative entropies. For example, the 
internal vibrations of the bases are likely to be very similar in the stacked
and unstacked states, and so the loss of these internal degrees of freedom is
likely to be of little thermodynamic consequence. 

Upon stacking, the bases lose entropy due to the more restricted nature of both
the relative positions of the centres of mass and the relative orientations of
the bases. If a model does not contain all these potential geometric sources of
entropy, for example, if, as many models do, a base is represented by a single
site with no orientational degrees of
freedom,\cite{Drukker00,Drukker01,Knotts07,Sambriski09,Sambriski09c,Kenward09,Tito10,Allen11,Freeman11,Linak11,Hsu12}
or if these degrees of freedom are not captured correctly, then it is likely
that an entropic term in the stacking interaction would be needed to
compensate.

Furthermore, most models also coarse-grain away the water, and the effects of
this on the thermodynamics is much harder to predict. Stacking is partly driven
by the hydrophobic character of the bases, but correctly describing 
hydrophobicity is far from straightforward and is known to often have a
strong entropic component.\cite{Chandler12} Therefore, it would
be unsurprising if an entropic component to the stacking would be required to
account for the contribution from hydrophobicity.

The second issue that needs careful consideration when designing a
coarse-grained model is the detailed forms used to describe the interactions,
with physically sensible choices of these forms being a prerequisite for a good
model. 
Here, we highlight what we think are key aspects of any model.

One important feature is the origin of the helicity of dsDNA and the
flexibility of ssDNA.  Of course, all models should have the double helical
B-DNA geometry as the global free-energy minimum of complementary strands
below their melting point.  One way to achieve this is to have the B-DNA
geometry as the minimum of all the individual interactions. However, this
choice may have significant consequences for the structure of ssDNA. 
In particular, this approach can lead to a backbone torsional potential that
imposes helicity on a single DNA strand, and hence to an unphysically rigid
helical geometry for ssDNA.\cite{Knotts07,Sambriski09,Sambriski09c} In fact,
DNA's helicity results from the combined constraints of the backbone and the
stacking interaction, and in particular, the difference in distance between
the separation of stacked bases ($3.4$\,\AA) and that typical for the
separation of bases along the DNA backbone (approximately $\sim 6.5$\,\AA),
with a (double) helix being the most favourable way of satisfying
both.\cite{Calladine} Thus, a model that achieves duplex helicity through
these two distance scales has the advantage that ssDNA when
unstacked is flexible and can kink sharply, although it can still also adopt a
helical geometry when stacked.

Interestingly, a recent coarse-grained model for dsDNA that was derived by
rigorous fitting to extensive all-atom simulations indicated that the
individual interactions are not all optimally satisfied in the B-DNA
structure.\cite{Gonzalez13} The ``frustration'' that results may be key to
explaining non-local effects of sequence on the double helical
structure,\cite{Gonzalez13} and may also be important to capture the
thermodynamic destabilization of the duplex that arises from
mismatches.\cite{Ouldridge11}

A second important feature is to capture the anisotropic nature of the
interactions between bases, both for stacking and hydrogen-bonding. As noted
already, if a base is represented by a single site without orientational
degrees of freedom, there is the potential problem that it will be hard to
capture the entropy loss associated with stacking. Furthermore, Watson-Crick
base-pairing interactions should be able to occur to one other base, when the
bases are coplanar and the chains are anti-parallel.  However, if the
description of the hydrogen-bond interaction between base sites is
isotropic,\cite{Knotts07,Sambriski09,Sambriski09c,Kenward09} there is the
possibility that the ``single-valent'' nature of Watson-Crick base pairing can
be broken. For example, this is exactly what Florescu and Joyeux found to occur
for the 3SPN model of de Pablo and
coworkers\cite{Knotts07,Sambriski09,Sambriski09c}  for a polydA-polydT duplex.
The two strands spontaneously slip by half a base-pair rise with respect to
each other so that each base can bind two bases on the complementary
strand.\cite{Florescu11} Note that this effect is normally prevented by the
heterogeneity of a sequence. 

In summary, although there are now a considerable number of coarse-grained DNA
models available, most are less-suited to studying the self-assembly processes
associated with DNA nanotechnology, because either they do not meet all the
requirements set out at the beginning of this section or their behaviour has
not yet been sufficiently well characterized. Of course, this is not to say
that these models do not have their own domains of applicability where they
can be productively used.  OxDNA, the coarse-grained model created in our
groups in Oxford, was developed with such nanotechnological applications
explicitly in mind, and it is our contention that it is currently the most
well-suited model for simulating DNA nanotechnology; it is also the model that
has been applied to the most DNA nanotechnology
systems.\cite{Ouldridge10,Ouldridge13,Sulc13} Therefore, for the rest of the
article we focus on this model.

Nevertheless, we note that if one is only interested in the structural
properties of DNA nanotechnology systems, other simulation approaches can
potentially be used.  For example, if the system is sufficiently small it may
be possible to use all-atom simulations, as has been done for a few
examples.\cite{Maiti04,Maiti06} Also for structures in which all the bases are
hybridized, coarse-grained models that describe the double-helical structure
well could potentially be applied, as long as the effects of junctions can be
incorporated into the model. For example, the rigid base-pair model of Ref.\
\citenum{Mergell03} has been adapted to probe the structure of DNA
origamis,\cite{Arbona12} and the CanDo package, which assumes that double
helices behave like elastic rods, is a useful resource for predicting DNA
origami structure and flexibility.\cite{Castro11,Kim12}

\section{The oxDNA model}
\label{sect:oxDNA}

\begin{figure}[t]
\centering
  \includegraphics[width=8.6cm]{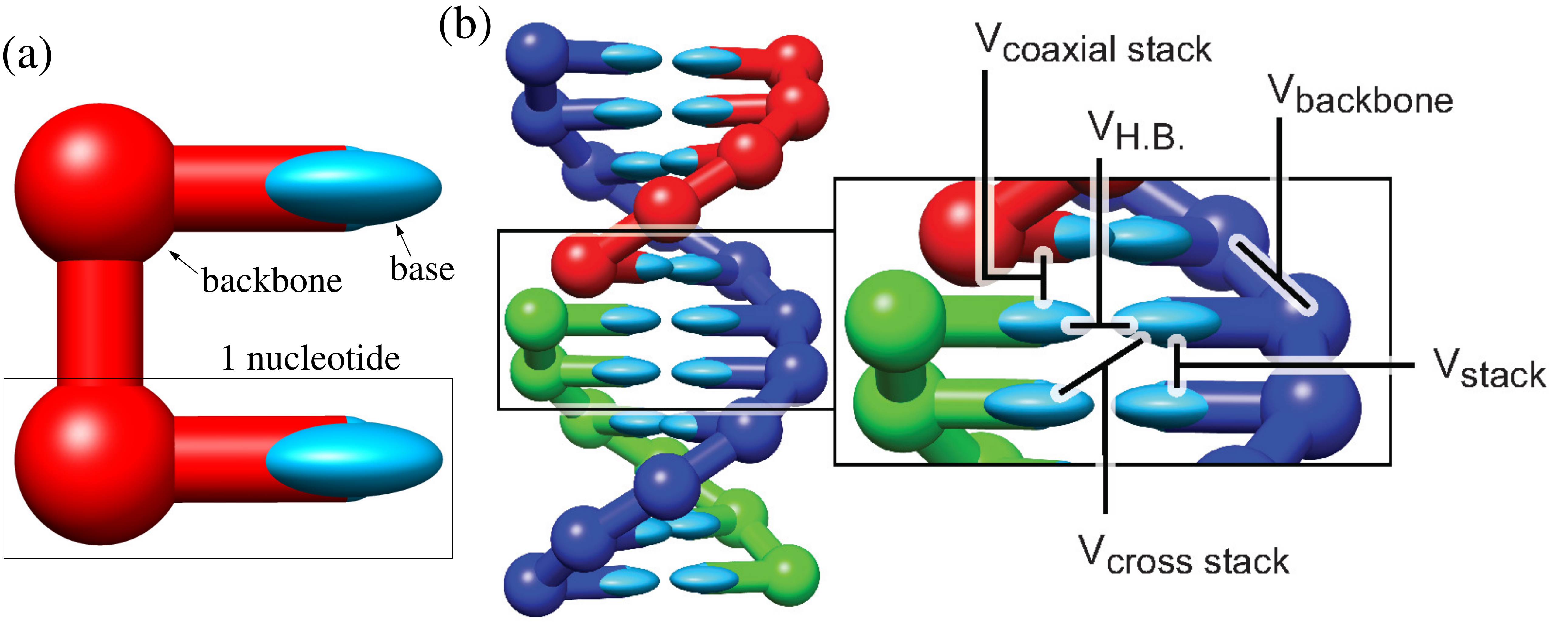}
  \caption{(a) A representation of the rigid nucleotides that are the
basic unit of the oxDNA coarse-grained model. The bases are represented by an
ellipsoid to reflect the orientational dependence of the interactions.
(b) Three strands in an 11-base-pair double helix with stabilising interactions
indicated in the inset. All nucleotides also interact through short-ranged 
excluded volume repulsions.
}
  \label{fig:basic}
\end{figure}

In the oxDNA model each nucleotide is represented by a rigid nucleotide that
consists of a set of collinear interaction sites and a vector that is
perpendicular to the notional plane of the base (see Fig.\ \ref{fig:basic}(a)).
The aim of the vector is to capture the planarity of the base through the
orientational dependence of the interactions rather than through additional
sites. This orientational dependence allows the model to represent the coplanar
base stacking and both the linearity of hydrogen-bonding and the edge-to-edge
character of the Watson-Crick base pairing. Furthermore, it is also used to
capture more subtle features of the interactions that are consequences of more
detailed structural features of DNA that are not present in the model, for
example, the right-handed character of the double helix and the anti-parallel
nature of the strands in the helix.

Note also that, in contrast to a number of models, there is a single
backbone site, rather than separate sites representing the sugar and phosphate
groups. We feel that this is a reasonable approximation for most of the
systems in which we are interested, but it may affect detailed structural
properties for systems where the backbones come in close proximity. 
One practical advantage of this choice is that it allows us to maintain the
pairwise nature of the internucleotide interactions without the need for
three-body terms to enforce the geometry of the backbone.  Furthermore,
coarse-graining the whole nucleotide into a single rigid body allows us to
ignore the internal motions of the nucleotides, thus significantly speeding
up sampling.

The form of the internucleotide potential is 
\begin{eqnarray} 
V_{\rm oxDNA} &= & \sum_{\left\langle ij \right\rangle} \left(
V_{\rm{b.b.}} + V_{\rm{stack}} + V'_{\rm{exc}} \right) + \nonumber \\ 
&& \sum_{i,j \notin {\left\langle ij \right\rangle}} \left( V_{\rm{HB}} +
V_{\rm{cr.st.}}  + V_{\rm{exc}}  + V_{\rm{cx.st.}} \right) ,
\label{eq:potential} 
\end{eqnarray} 
where the first sum is taken over all nucleotides that are adjacent along the
backbone of a strand and the second sum is taken over all other pairs of
nucleotides. The different terms represent backbone connectivity
($V_{\rm{b.b.}}$), excluded volume ($V_{\rm{exc}}$ and $V'_{\rm{exc}}$),
hydrogen bonding between complementary bases ($V_{\rm{HB}}$), stacking between
adjacent bases on a strand ($V_{\rm{stack}}$), cross-stacking
($V_{\rm{cr.st.}}$)  and coaxial stacking ($V_{\rm{cx.st.}}$).  The
nucleotides that participate in these interactions within the double-helical
state are schematically shown in Figure \ref{fig:basic}(b). The excluded
volume and backbone interactions are isotropic, whereas all other interactions
depend on the relative orientations of the nucleotides as well as the distance
between the interaction sites.  The detailed forms for all of the terms in
Eq.\ \ref{eq:potential} are available in Ref.\ \citenum{Ouldridge11b}.  Note
that these forms are not attempting to provide a chemically realistic
description of the interactions, but instead a physically realistic
description of effects of the interactions.

We applied a top-down approach to parameterize the oxDNA model.  Loosely
speaking, the structure is determined by the positions of the minima in the
interaction potentials, the thermodynamics by the well depths and force
constants, and the mechanical properties also by the force constants, and all
three types of data have been used to fit the parameters of the model.  In
particular, for the thermodynamics, we have fitted the model to reproduce the
transition temperatures and widths for the melting of short duplexes as
predicted by the SantaLucia model,\cite{SantaLucia04} and results for the
stacking transition of ssDNA.\cite{Holbrook99}
Similarly, for the mechanics we have fitted to the persistence lengths of
ssDNA and dsDNA and the elastic moduli of dsDNA. We note that for the fitting
of the model it would have been very useful to have more unequivocal
experimental data on the physical properties of single-stranded DNA,
particularly for the stacking transition.

As noted in the previous section, the effective interactions in a
coarse-grained model should in principle be free energies.  However, if the
coarse-graining maintains the correct relative entropies between the states of
interest, the entropic components of the interactions would be zero.  As we
have tried to retain all the geometric sources of entropy relevant to
hybridization (e.g.\ the orientational degrees of freedom of the bases and the
flexibility of the single strand), we attempted to keep the interactions
temperature-independent as far as possible. In the end, it only proved
necessary to introduce temperature dependence into the stacking interactions
(and only to a relatively small degree) in order to obtain a good fit to the
thermodynamic data. This suggests that our model has been relatively
successful in retaining the relevant geometric degrees of freedom; of course,
we are unable to retain the entropic components of the interactions related to
the solvent, and so the need to introduce temperature-dependence into oxDNA's
stacking interactions,
which have a significant hydrophobic component, might reflect this source of
entropy. 

There are currently two parameterizations of the oxDNA model available.
Firstly, there is an ``average-base'' parameterization in which the
interactions for the different bases are identical except for the
hydrogen-bonding term, for which bonding only occurs between Watson-Crick base
pairs.\cite{Ouldridge10,Ouldridge11} In the second parameterization, the
interaction strengths for stacking and hydrogen-bonding have been allowed to
vary in order to capture the sequence-dependence of DNA
thermodynamics.\cite{Sulc12} However, the development of this
sequence-dependent version has not made the average-base model redundant,
since it is often useful to examine the generic behaviour of DNA, unobscured
by sequence-dependent effects. Indeed, most of the published applications of
oxDNA have been reported for the average-base
parameterization.\cite{Ouldridge10,Romano12b,Matek12,Romano13,Ouldridge13,Sulc13,Srinivas13,Ouldridge13b,deMichele12}
By contrast, the sequence-dependent parameterization is particularly useful
when we want to compare in detail to a particular experimental system.

As with any coarse-grained model, it is important to be open about oxDNA's
limitations.  Firstly, the model is parameterized for a specific salt
concentration, namely 500\,mM, and the electrostatics interactions between
charged groups are not explicitly represented but instead their effects are
incorporated into the excluded-volume interactions. This approach is
reasonable because of the short-range nature of the Debye screening length at
this salt concentration. Furthermore, these solution conditions are not
untypical of the relatively high ionic strengths used in DNA nanotechnology
experiments.  
In principle oxDNA could be reparameterised at different salt concentrations,
but at lower salt it is not clear the short-ranged functional forms we employ
would be sufficiently flexible. Some progress could perhaps be made by through
a Debye-H\"{u}ckel like description of the electrostatics, as has been done
for a number of other DNA models.\cite{Knotts07,Sambriski09,Sambriski09c,MorrisAndrews10,Hsu12,Edens12} However, it
should be kept in mind that further complications such as non-linear
electrostatics also become increasingly important at low salt, see e.g.\ Refs.
\citenum{Kornyshev07} and \citenum{Mocci12} for further discussion.  

Secondly, even for the second parameterization of the oxDNA potential, the
sequence-dependence is limited. All properties of the bases are assumed to be
the same except for the strengths of the attractive interactions; for example,
all bases are still of the same size. For this reason, the model is unlikely 
to be able to address the sequence dependence of the detailed structure and
elasticity of dsDNA. 

Thirdly, the double helix in oxDNA is symmetrical with equal sizes for the
major and minor grooves (Fig.\ \ref{fig:basic}(b)). For many applications this
deficiency will be of little importance, but it will affect systems where the
detailed geometry of the backbone is significant.  For example, major and minor
grooving is needed to reproduce the correct relative angles between crossovers
in DNA origami when both staple and scaffold strand crossovers are present.
For this reason, we are in the process of parameterizing a version of the
model that includes the correct grooving.

We note that simulation codes incorporating the oxDNA potentials and
all the algorithms presented in the following section are available
from the oxDNA web-site.\cite{oxDNA}

\section{Simulating coarse-grained models} 
\label{sect:sims}

DNA molecules in solution undergo diffusive motion. In molecular dynamics
simulations of all-atom models, this diffusional dynamics is achieved
through collisions between the DNA and the solvating water molecules, which also act as a thermal and pressure reservoir. However, in virtually all
 coarse-grained DNA models, the solvent is modelled implicitly and so applying
standard molecular dynamics to such models would generate artefactual ballistic
dynamics for the DNA. Further, simulations would be isoenergetic, whereas sampling from the canonical ensemble is appropriate when models of dilute systems with implicit solvent are to be compared directly to experiment.\cite{Ouldridge12}

A solution to this problem is to use Langevin dynamics,\cite{Snook07} in which additional drag and
random forces are added to Newton's equations of motion in order to achieve
diffusional dynamics. The size and form of these forces can be chosen self-consistently so that the system samples from the canonical ensemble. In the limit of strong noise and frictional forces, inertia can be neglected in a formalism known as Brownian dynamics. A simpler approach is to use Andersen-like thermostats,\cite{FrenkelSmit2} in which systems are evolved using Newton's equations between occasional `collision steps' when particle momenta are resampled from the thermal distribution. On scales longer than the typical time for resampling momenta, motion of particles is diffusive (although it is ballistic on shorter times). 

Langevin, Brownian and Andersen-like thermostats all produce diffusive motion
and sample from the canonical ensemble. In most implementations, however, they
neglect long-range hydrodynamic forces which (amongst other effects)
accelerate the diffusion of large clusters of particles. As a consequence,
relative diffusion rates of different size clusters will not scale correctly
when systems are simulated using these approaches. It is possible to
incorporate long-range hydrodynamic effects in Langevin or Brownian
algorithms,\cite{Snook07} but generally this is not done due to the extra
computational cost.

Issues with cooperative hydrodynamics highlight the problem of comparing to
experimental time scales. Designing a model with mass, energy and length
scales then automatically defines a unit of time, which in principle could be
used to compare with experiment. However, many experimental measurements
reflect processes which require overall diffusion of strands, and as we have
pointed out most dynamical implementations do not reproduce the physical
scaling of diffusion coefficients with system size. An added complication is
that coarse-grained models tend to smooth out the microscopic roughness of
energy landscapes, thereby accelerating motion.\cite{Hills10}  Given these
caveats, it seems sensible to focus primarily  on relative rates of similar
processes when simulating coarse-grained models, where uncertainties in
mapping time scales will tend to cancel out. We note that the point of
coarse-graining is to make simulations faster, and so one can even
deliberately accelerate dynamics by using artificially high diffusion
coefficients. For oxDNA at least, we have shown that, as long as one is
careful to compare relative rates, such an approach has no qualitative
consequences for the process of duplex formation.\cite{Ouldridge13b} More
generally, we expect this approach to be be reasonable if there is a
separation of the time scales between diffusion and the relevant internal
rearrangements of the strands. Indeed, an alternative way to map time
scales onto experiment is to simply scale the simulation time to match the 
experimental diffusion constant of a strand, although this approach becomes
more complicated when the simulations involve strands of different length 
because coarse-grained simulations do not generally reproduce the physical 
scaling of the diffusion coefficient with size.

Monte Carlo is essentially a method to perform equilibrium sampling.
However, if the aim is simply to reproduce diffusive motion and sample from the
canonical ensemble, Monte Carlo methods using local moves can provide an
alternative approach to simulate
dynamics.\cite{Kikuchi91,Berthier07,Tiana07,Sanz10}
However, if one were to  apply a Monte Carlo algorithm that used ``single
particle'' moves to oxDNA, the simulations would be quite inefficient, because
a significant fraction of the moves that involve strongly interacting
nucleotides would be rejected.  As a result, the diffusion of strands would be
particularly slow, and even more so if the strands were part of duplexes or
higher-order aggregates. Thus, single-particle move Monte Carlo will 
drastically fail to reproduce the relative diffusion coefficients of different-sized aggregates.

One way to overcome this problem is to use a Monte Carlo algorithm that
attempts moves of clusters of nucleotides.  One such algorithm that we have
found to be particularly useful is the Virtual Move Monte Carlo (VMMC)
algorithm of Whitelam and Geissler\cite{Whitelam07} (note that we actually use
the variant in the appendix of Ref.\ \citenum{Whitelam09b}). One of the
advantages of this algorithm is that the clusters that are constructed reflect
both the current configuration and the proposed move.
As a result, the algorithm  allows more efficient sampling for our DNA model
than Monte Carlo with single-nucleotide moves. We note here that  cluster
algorithms work best when the system is dominated by pairwise potentials, as
multi-body interactions cannot generally be used in the cluster growth
procedure.

In principle, one can modulate the VMMC algorithm to try and capture relative
diffusion rates of clusters of different sizes.\cite{Whitelam07} Reproducing
physical diffusion rates is far from simple, however, and may have consequences
for efficiency as the techniques proposed involve rejecting moves that would
otherwise be accepted. It is also difficult to directly compare simulation
times for systems of different sizes, or when an external force is applied in
one case and not in another. Nonetheless, we have found that VMMC trajectories
often give qualitatively similar results to Langevin and Andersen-like
thermostats for oxDNA, suggesting that the kinetic results are not artefacts of
a particular choice of algorithm.  One possible approach to generate dynamical
Monte Carlo trajectories would be to combine VMMC with simpler cluster
algorithms, such as that of Bhattacharyay and Troisi.\cite{Bhattacharyay08}
VMMC could be used to capture internal relaxation of clusters and the simpler
algorithm to drive overall diffusion.  

In our simulations of oxDNA, we have used a variety of the above techniques.
VMMC has proved to be extremely computationally efficient for simulating small
systems (smaller than about 200 nucleotides) in which processes involving
hybridisation or melting of duplexes are of interest. It is also easy to
couple to rare-event methods such as umbrella sampling (see section
\ref{rare-event methods}), making it ideal for measuring the thermodynamics of
DNA reactions.  The efficiency of VMMC stems from the ability to make
relatively large moves without losing precision, unlike methods that involve
integrating equations of motion. We note that very large moves that could
cause strands to pass through each other must be excluded, however, if
topology preservation is required.  

In general, we have used Langevin\cite{Davidchack09} and
Andersen-like\cite{Russo09} algorithms (specifically designed for rigid bodies)
to study reaction kinetics, as the dynamics is easier to interpret than for
VMMC (even if more computational time is required to observe processes). For
equilibrating structural properties of large systems such as DNA origamis or
long duplexes, we have used the Andersen-like algorithm as it is most
computationally efficient (VMMC  is less effective at performing internal
mechanical relaxation for large clusters, and generates many unhelpful moves of
the entire system). Additionally, Langevin and Andersen-like thermostats are
easier to parallelize than cluster-building Monte Carlo approaches, which are
naturally more serial in character. Implementing the Andersen-like algorithm on
graphical processing units has allowed us to treat extremely large systems of
tens of thousands of nucleotides, and so sample the structural equilibrium of
DNA origamis and reactions involving many strands.

\subsection{Rare-event methods}
\label{rare-event methods}
Although coarse-grained models allow much longer time scales to be simulated
than for all-atom models, many of the processes involving DNA in which one
might be interested involve significant free-energy barriers, and so may still
be hard to sample. For example, free-energy barriers can arise due to the loss
of translational entropy when two species associate, or due to the necessity of
breaking base pairs, or if a process is geometrically or topologically
constrained.\cite{Romano12b} In characterising such processes, rare-event
simulation techniques are extremely useful; we shall first consider techniques
designed to measure the equilibrium properties of such systems, before dealing
with approaches for measuring kinetics.

\subsubsection{Accelerating equilibration.} 

Two methods of particular note for improving equilibration are umbrella
sampling\cite{Torrie77,FrenkelSmit2} and parallel tempering.\cite{Earl05}
Umbrella sampling enhances the equilibration of thermodynamic properties by
artificially biasing the formation of intermediate configurations between two or
more local free-energy minima, thereby increasing the rate of transitions
between these minima. The sampling bias can then be corrected to yield an
unbiased free-energy landscape for the process in question. Parallel tempering
involves running  simultaneous replica simulations at a range of temperatures,
and occasionally swapping configurations between replicas with a probability
that preserves the canonical distribution at each temperature. This approach
can accelerate equilibration if reactions are faster at temperatures above or
below the one in question. In general, we have found the precise control
offered by umbrella sampling to be most useful in measuring thermodynamic
properties of our systems, with parallel tempering an alternative when it is
difficult to define an effective order parameter.\cite{Romano13} 

When applying such rare-event methods to the formation of a target structure
from more than one species, it is usually most computationally convenient to
consider the formation of a single target, e.g.\ one DNA duplex from two
complementary strands. However, the statistical distribution of various
clusters of strands within a simulation volume obtained from a single-target
simulation is different from the distribution that would be obtained from a
bulk simulation of the same model (at the same concentration and
temperature).\cite{Ouldridge10b,Ouldridge12} Physically, the difference arises
because concentration fluctuations are suppressed. For example, in the case of
a single duplex, there would always be one strand of each type in the simulated
volume, whereas for the same volume in a bulk system there are many other
possibilities. Assuming that the concentration of duplexes in the simulated
system is representative of bulk leads to a melting transition that is much
narrower (as temperature is varied) than would be observed in bulk (the width
is approximately 50\% larger in bulk). For non-self complementary duplexes, the
duplex concentration is also systematically overestimated at all temperatures,
leading to an error in the melting temperature of several Kelvin, depending on
the width of the transition. To avoid this problem, one could of course choose
to simulate a larger system that is able to form multiple copies of the target
structure.  However, the finite-size effects associated with suppressed
concentration fluctuations can persist up to surprisingly large system sizes,
particularly for assemblies involving many strands.\cite{Ouldridge10b}
Furthermore, it is often no longer straightforward to apply rare-event
approaches
because of the problem of defining an appropriate order parameter to drive the
formation of multiple copies of a target, or the need to use more replicas for
parallel tempering.

We note that, to date, these finite-size effects have generally been
neglected.\cite{Sambriski09,Prytkova10,Tito10,Freeman11,Allen11,Hoefert11,Schmitt11,He13}
However, we have recently developed an analytic approach
that allows the bulk probabilities to be estimated from the ``single-target''
simulations assuming both that the species behave ideally and that other
multi-chain aggregated states do not contribute to the
thermodynamics.\cite{Ouldridge10b,DeJong11,Ouldridge12} Both assumptions are
likely to hold well for DNA, firstly, because ideality is a good approximation
for the dilute DNA solutions typically used, and secondly, because the targets
one usually wants to consider are maximally base-paired, so any alternative
misbonded aggregate, although perhaps relevant as a kinetic trap, is likely to
have higher free energy than the target and so not be thermodynamically
relevant.  This procedure is simple to implement, and so we recommend that it
should be applied for any calculation of thermodynamic properties from
single-target simulations. 

\subsubsection{Accelerating kinetic measurements.}

Accelerating kinetic measurements is more difficult than enhancing
thermodynamic equilibration, as simulations must be unbiased and transitions
must occur at the temperature of interest. In general, methods deal with
measuring the flux of trajectories from one local free-energy minimum $A$ to an
alternative $B$ (a quantity analogous to transition rates for non-instantaneous
processes), and also generating a representative set of transition paths.

We have found forward flux sampling to be a particularly useful algorithm of
this kind.\cite{Allen05,Allen2009} First, the transition is split into several
stages by defining interfaces which must be crossed in the transition from $A$
to $B$. Simulations are performed in the vicinity of $A$ to measure the flux of
trajectories across the first interface. States are saved at the crossing
point. These states are then used to initiate new trajectories, which are used
to measure the probability of reaching the next interface before returning to
$A$. This process is repeated for subsequent interfaces.The overall flux from
$A$ to $B$ is then given by the initial flux multiplied by the probability of
success at each interface, and the trajectories generated are an unbiased
sample from the ensemble of reaction pathways. By splitting one unlikely
process into several stages, the overall measurement can be performed more
efficiently.  

Forward flux sampling is most effective for simple processes which do not
involve long-lived metastable intermediates, such as hybridisation of
non-repetitive duplexes.\cite{Ouldridge13b} Although the algorithm is still
valid when such metastable intermediates exist, it does not effectively
accelerate the resolution of the intermediate into state $A$ or $B$, and hence
this subprocess must be simulated by brute force. In some cases this is
practical, but often it is not. This difficulty arises because one must measure
the number of trajectories that return to state $A$ at each stage, rather than
just the previous interface. 

A possible solution to this problem is to explicitly define a state $C$
corresponding to the metastable intermediate. One can then use forward flux
sampling to measure the flux from $A$ to $C$, and then from $C$ to $B$ and $A$,
from which the overall kinetics of the process can be inferred. It is also
possible to consider multiple metastable intermediates. Such an approach, which
we have used to study internal displacement of repetitive
sequences,\cite{Ouldridge13b} relies on the assumption that the system
equilibrates in state $C$ before it resolves into state $A$ or $B$. We note
that alternative techniques, such as transition path sampling\cite{Dellago02}
and transition interface sampling,\cite{vanErp05} face similar difficulties
when confronted with metastable intermediates.

\begin{figure}[t]
\centering
  \includegraphics[width=8.6cm]{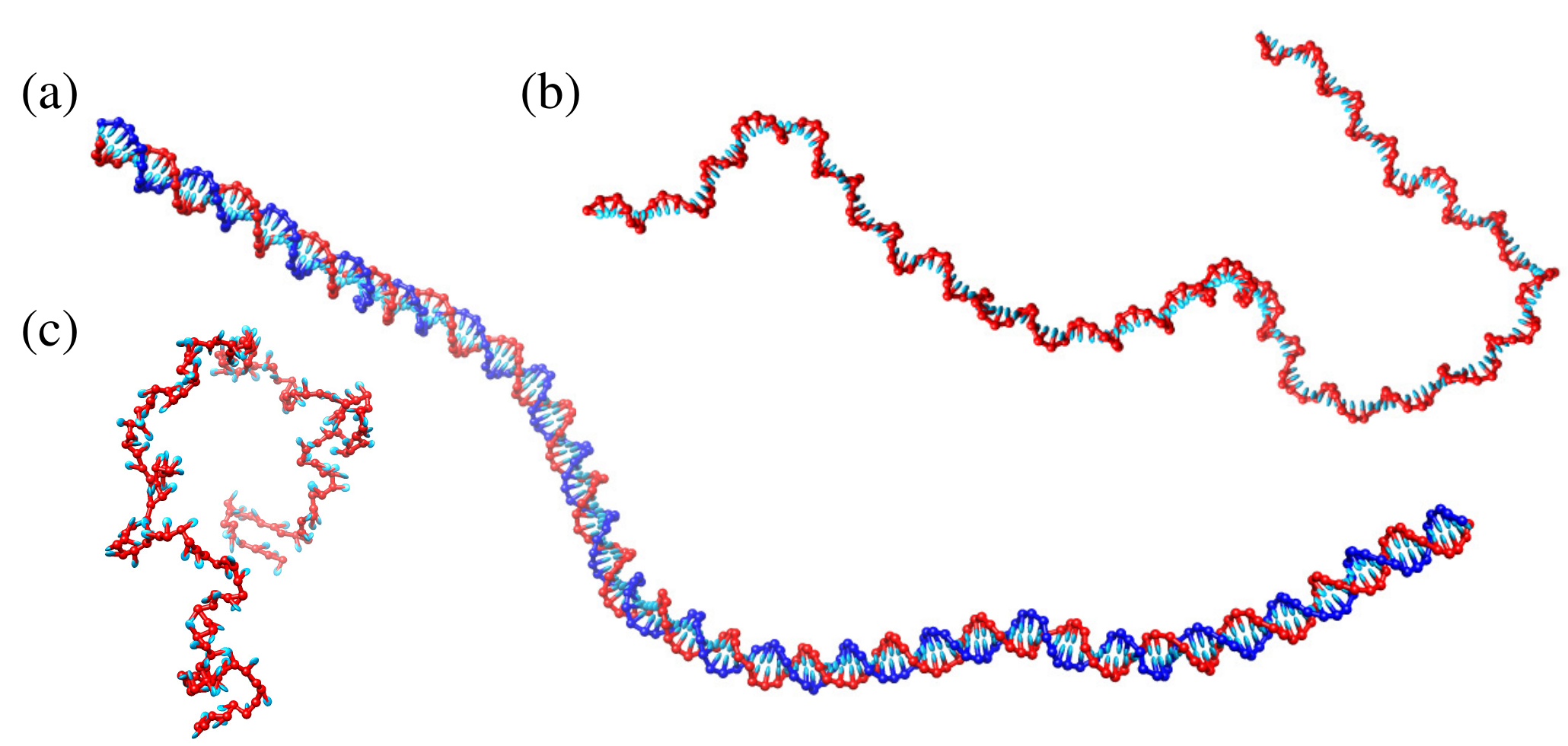}
  \caption{Typical configurations illustrating the relative flexibility of 
(a) double-stranded DNA, (b) stacked single-stranded DNA and (c) unstacked
single-stranded DNA. }
  \label{fig:persistence}
\end{figure}

\section{Results for oxDNA}
\label{sect:results}

\subsection{Basic DNA biophysics}

Fig.\ \ref{fig:basic}(b) shows a DNA double helix as represented by oxDNA.  By
design it has the correct basic structure, e.g.\ pitch, base-pair rise and
radius. However, as already noted, the two grooves are of the same size,
whereas in real DNA the major groove is larger. 
The double helix also exhibits non-trivial features such as propeller twist of
the bases that arises in the model due to a competition between optimizing the
stacking and hydrogen-bond interactions.\cite{Ouldridge11}

\begin{figure*}[t]
\centering
  \includegraphics[height=8cm]{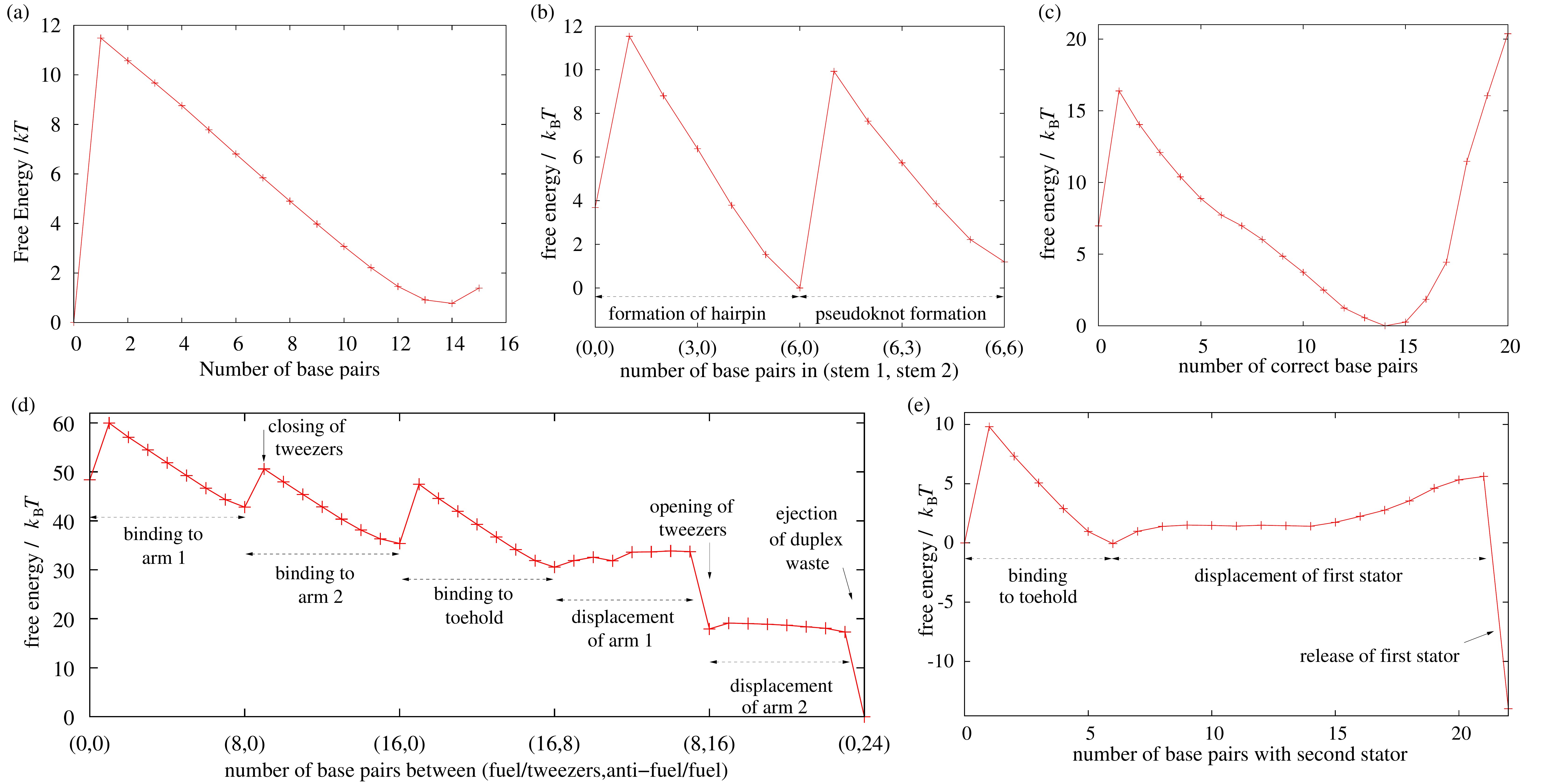}
  \caption{Free-energy profiles for (a) the formation of a 15 base-pair duplex
at 343\,K 
(b) the formation of a pseudoknot at 307\,K 
(Fig.\ \ref{fig:structures_intermediate}(d)),
(c) the formation of a kissing complex at 296\,K
between two hairpins with 20-base complementary loops
(Fig.\ \ref{fig:structures_intermediate}(e)),  
(d) the complete cycle of the DNA nanotweezers illustrated 
in Fig.\ \ref{fig:nanodevices}(a),  
and (e) one step of the burnt-bridges walker illustrated 
in Fig.\ \ref{fig:nanodevices}(b).  
All the results are for the average-base parameterization of oxDNA.
In (c) the topological frustration
leads to a significantly lower free-energy gain from binding than for
a duplex with the same number of bases and the most favourable state of the
kissing complex has only 14 of the 20 possible base pairs formed.
}
  \label{fig:FEL}
\end{figure*}

The model also provides a good description of the mechanical properties of DNA.
For example, Fig.\ 2 illustrates the relative persistence lengths of dsDNA and
ssDNA, in its stacked and unstacked forms, which in our model are 125 base
pairs, and 40 and 3 bases, respectively, in good agreement
with available experimental data.\cite{Ouldridge11}  In particular, the relative
flexibility of unstacked DNA is important for nanotechnological applications.
This flexibility also enables hairpins to form easily.

Perhaps unsurprisingly, given their importance in the fitting procedure, we
are able to reproduce well the melting points and transtion widths of DNA
duplexes. In order to capture the dependence of the melting point on length,
the correct balance between the contributions of stacking and hydrogen-bonding
to the stability of the duplex was required. 
More impressively, the melting points of hairpins are also reproduced well,
albeit with a small but constant offset of approximately 2\,K. Single-strands
in oxDNA also show a broad uncooperative stacking transition, although in this
case the experimental nature of this transition is less clear. However, some
further information on this transition can be obtained indirectly from the
melting temperatures of hairpins, which decrease when the stacking in the loop
is sufficiently strong to hinder the bending back of the strand in the
loop,\cite{Goddard00} an effect that our model is able to
reproduce.\cite{Sulc12}

One of the features of our model is that it allows the free-energy landscapes
associated with various assembly processes to be
computed.\cite{Ouldridge10,Ouldridge11,Romano12b,Matek12,Sulc12,Romano13,Ouldridge13,Sulc13,Srinivas13}
For example, Fig.\ \ref{fig:FEL}(a) depicts the free-energy profile for the
hybridization of a 15-base-pair duplex, where the number of base pairs has
been used as the order parameter. There is an initial free-energy barrier
associated with the loss of translational entropy on association followed by a
linear downhill slope as the number of base pairs increase (the linearity is
because the free-energy of forming a base pair is the same for all base pairs
in our ``average base'' parameterization). Notably, the most stable state of
this duplex at this temperature is with only 14 of the 15 bases paired. This
entropic opening of the duplex at its ends is termed ``fraying'', and provides
an example of how oxDNA's behaviour shows deviations from ``two-state''
behaviour; this contrasts with, for example, SantaLucia's nearest-neighbour
model for predicting oligonucleotide melting temperatures that assumes the two
states in equilibrium are a fully-formed duplex and an unstructured single
strand with no variation of the states involved with
temperature.\cite{SantaLucia04}

A simplistic interpretation of this free-energy profile might be to say that
the ``transition state'' to duplex formation involves the formation of a single
base pair. However, it is not quite that straightforward, because even when a
single base pair forms the rest of both chains normally require considerable
conformational change to reach a state where the duplex can zipper up, and so
consequently the system is actually much more likely to
dissociate.\cite{Ouldridge13b}  Similarly, when a duplex is melting and the
last base pair breaks, it does not initially feel the extra translational
entropy it can gain from being disassociated, and so is still quite likely to
zipper back up.

\begin{figure}[t]
\centering
  \includegraphics[width=8.6cm]{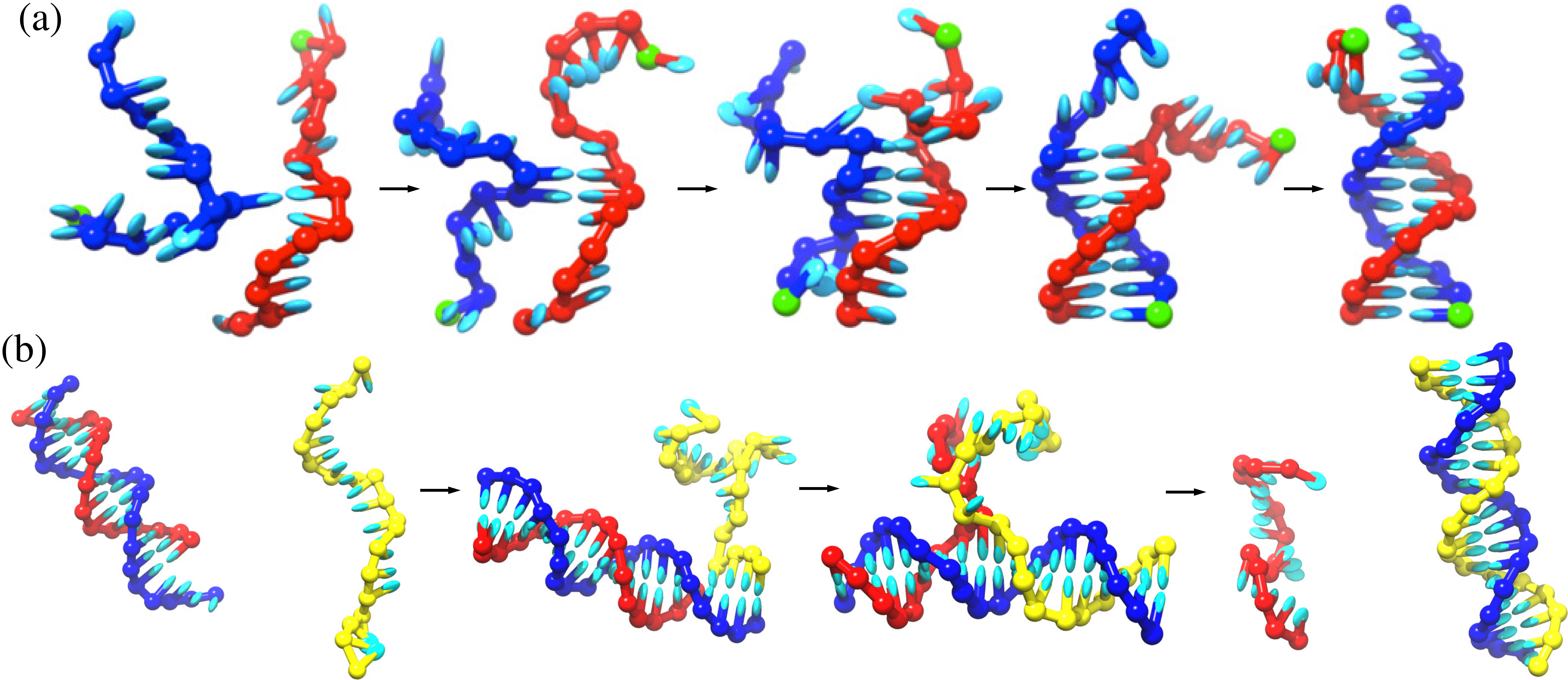}
  \caption{Example configurations from the pathways for 
(a) hybridization 
and 
(b) toehold-mediated strand displacement. 
}
  \label{fig:basic_dynamics}
\end{figure}

We should note that the mechanism of hybridization that we observe for oxDNA
differs strongly from that found by de Pablo and coworkers for their 3SPN.1 DNA
model.\cite{Sambriski09,Sambriski09b,Sambriski09c,Schmitt11,Schmitt13} For
oxDNA, hybridization normally occurs by the formation of a nucleus of a few
correct base pairs, followed by the ``zippering'' up of the rest of the duplex
as bases are transferred from the relatively unstructured single-stranded tails
to the growing end of the double-helical section, as illustrated in Fig.\
\ref{fig:basic_dynamics}(a).  Alternatively, if a misbonded (i.e.\ involving
base pairs not present in the fully-bonded duplex) nucleus forms first, the
fully base-paired duplex can still be achieved by an internal displacement
mechanism,\cite{Ouldridge13b} in which the correctly-bonded helical section
nucleates and grows at the expense of the misbonded helix. By contrast, for the 3SPN.1 model
ssDNA adopts an overly stiff helical geometry, and hybridization can occur from
a misbonded duplex by a sliding of the two chains past each other along the
helical axis. In our opinion,\cite{Ouldridge13b}  this mechanism is an artefact
of their overly stiff ssDNA that restricts the way DNA can hybridize.  In
particular, both the zippering mechanism shown in Fig.\
\ref{fig:basic_dynamics}(a) and the internal displacement mechanism would not
be feasible in 3SPN.1.  Furthermore, the isotropic nature of the 3SPN.1
base-base attractions and the {\em ad hoc} sugar-sugar attraction leads to an
artificially low barrier to sliding.

\begin{figure}[t] \centering
\includegraphics[width=8.4cm]{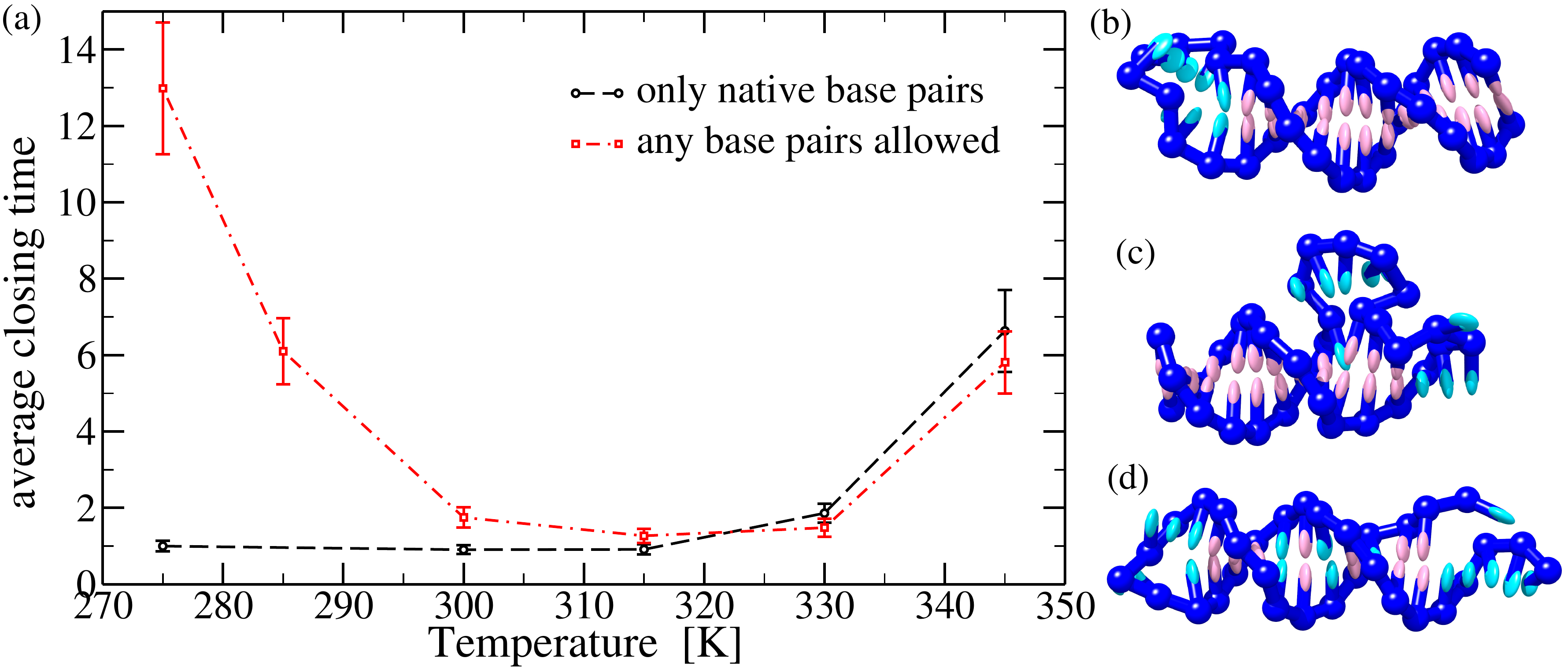} \caption{(a) Folding
times for a hairpin 
as a function of temperature for the sequence
TTACATAAAAAGTTTTTTTTCTTTTTATGTAA. 
Also represented are folding times when misbonding is prevented by only
allowing base pairs in the target hairpin to form.  The folding times are
given relative to the fastest folding system.
(b) The target hairpin. It has a 12-base-pair stem and an 8-base loop, and a
melting point of 347\,K.  (c) The most stable misbonded structure. It
has 11 base pairs and a melting temperature of 334\,K.  (d) The second most
stable misbonded structure, which has 7 base pairs and a melting temperature
of 314\,K. 
In (b)--(d) the base sites of nucleotides participating in Watson-Crick base
pairs are coloured pink to aid visualization of the bonding pattern.  }
\label{fig:hairpin_closing} \end{figure}

To illustrate some of the general principles of self-assembly in DNA systems,
in Figure \ref{fig:hairpin_closing}(a) we show the hairpin closing times for a
sequence that can form misbonded structures which can compete with the
formation of the correct hairpin. Like many other self-assembly processes,
there is a temperature window where assembly is most efficient. At high
temperatures, the hairpin formation rate goes down due to the decreased
probability of successfully forming a complete stem after the formation of an
initial contact. Furthermore, above the melting temperature, the equilibrium
fraction of hairpins will decrease rapidly. At low temperatures the hairpin
folding times go up due to the formation of misbonded hairpin structures that
act as kinetic traps and hinder the formation of the target hairpin. The
system can still escape from these misbonded structures by internal
displacement processes, but this becomes increasingly difficult as the
temperature is decreased. The two most stable misbonded structures are
illustrated in Fig.\ \ref{fig:hairpin_closing}(c) and (d). As it happens the
more thermodynamically stable of the two is a less effective kinetic trap,
because it can rearrange to form the target hairpin relatively easily just by
the propagation of the bulge loop to the hairpin loop end of the helical stem. 
If the formation of misbonded structures is prevented by artificially turning
off non-native base pairing, there is no low temperature decrease in the
hairpin formation rate.

This example shows some of the typical features of DNA self-assembly.  Namely,
the existence of an assembly window between $T_m$ (the temperature at
which the target structure melts) and $T_\mathrm{misbond}$ (the temperature at
which the most stable misbonded structure melts), where the only structure that
is stable with respect to the single-stranded state is the target structure.
This is why annealing the system, i.e.\ cooling it down from a high temperature
where all the strands are in the single-stranded state, is such a simple and
effective assembly strategy. The system will always pass through the assembly
window first and, as long the cooling rate through this window is sufficiently
slow, the correct structure will form and the system will not ever get
kinetically trapped in misbonded structures.  Sequence design can be used to
make self-assembly easier by minimizing the stability of misbonded structures,
hence making the assembly window wider.  In addition, for complex structures
that are designed to form hierarchically (i.e.\ subunits that form at high
temperature themselves assemble into larger aggregates at lower temperature)
there will be multiple assembly windows, which will be sequentially passed
through on annealing.

\subsection{DNA under stress}

In the cell, DNA is acted upon by a great variety of molecules, and many of
these induce some stress on the DNA.\cite{Cozzarelli06,Fogg12} For example,
protein binding can cause DNA to bend sharply (e.g.\ the wrapping of DNA around
histones in order to package DNA in eukaryotic cells).  There are also families
of enzymes responsible for manipulating DNA in different ways; e.g.\
topoisomerases control the supercoiling of DNA, and helicases cause DNA to unwind.

\begin{figure*}[t]
\centering
  \includegraphics[width=17.8cm]{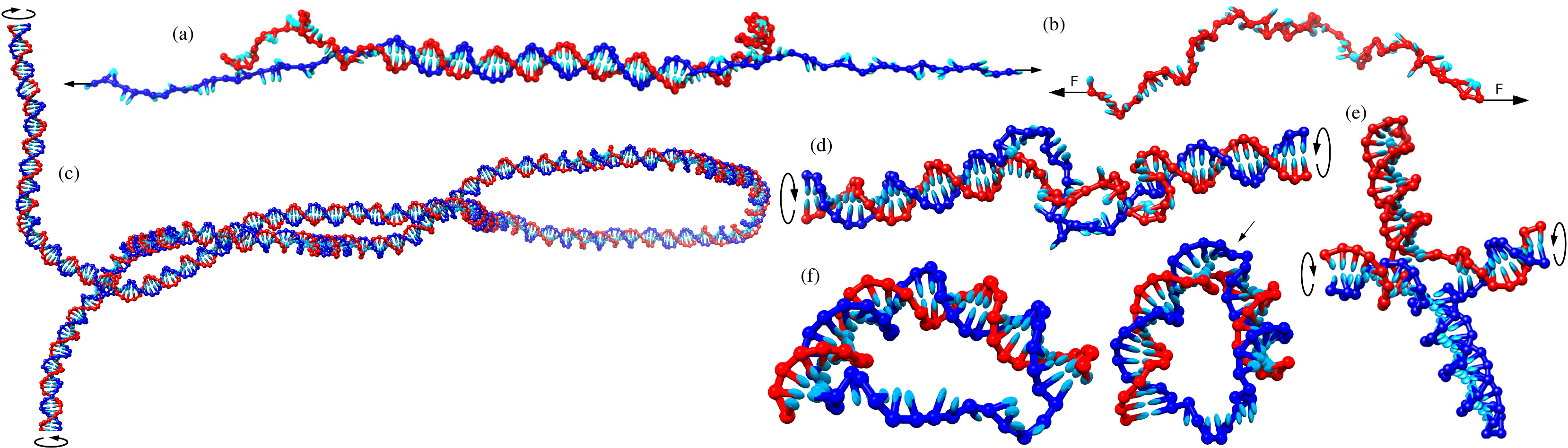}
  \caption{Response of DNA to mechanical stress. (a) The unpeeling of duplex
DNA and (b) the stretching of ssDNA under tension, and the formation of (c) a
plectoneme, (d) a bubble and (e) a cruciform for a sequence with an inverted
repeat, all under negative supercoiling. 
(f) Two structures formed by the hybridization of a 35-base linear strand with
a 45-base circular strand. For the left-hand structure, the double helical
section is homogeneously bent, whereas the other contains a ``kink'' (labelled
by an arrow) where two base pairs are broken.
}
  \label{fig:stress}
\end{figure*}

As such, the mechanical properties of DNA are of considerable
interest, and the advent of single-molecule techniques, such as optical and
magnetic tweezers, to exert forces and torques on individual DNA molecules has
led to a wealth of new and impressive data.\cite{Bryant12} However, such
techniques cannot directly detect the microscopic nature of the structural
changes induced by the mechanical stress, and so simulations can play an
important role in visualizing such structural changes.  As noted already,
oxDNA was fitted to give a reasonable description of the persistence length of
dsDNA. 
Similarly, the twist modulus of dsDNA 475\,fJ\,fm is at the upper end of the
range of values determined from experiment, although this comparison is
complicated by the fact that experiments often measure an effective twist
modulus that is reduced by thermal bending fluctuations.\cite{Moroz97}
However, the stretch modulus of 2100\,pN is roughly twice the experimental
value.\cite{Ouldridge11}  Note, when fitting the oxDNA model, parameters could
not be found that allowed the model to simultaneously reproduce the
experimental persistence length and stretch modulus and so we choose for the
model to better reproduce the persistence length, as this is more likely to be
relevant to the nanotechnological applications.

On pulling on opposite ends of dsDNA, it is found to undergo an
overstretching transition to an overstretched form with an approximately 70\%
increase in length. At room temperature and 500\,mM salt this occurs at
67--8\,pN.\cite{Wenner02,Zhang12} OxDNA undergoes a similar transition at
74\,pN, satisfyingly close to the experimental value, and also provides a good
description of the temperature dependence of the overstretching
force.\cite{Romano13} The nature of the overstretched state has been the
subject of considerable debate, the two main proposals being force-induced
melting by unpeeling\cite{Williams02b} and a transition to an overstretched
duplex state termed ``S-DNA'',\cite{Smith96} but with recent evidence
suggesting that both mechanisms can occur depending on solution conditions and
temperature.\cite{King13,Zhang13} Interestingly, for our model we only ever
see overstretching by unpeeling (Fig.\ \ref{fig:stress}(a)). This result
strongly suggests that S-DNA is unlikely to be an unstacked but base-paired
duplex, one of the proposed structures for S-DNA,\cite{Smith96} as our model
has a good representation of stacking and base pairing and would be expected
to reproduce such a structure if it were relevant to overstretching.  Instead,
S-DNA is likely to have a more exotic structure.

The stretching of ssDNA has been less studied, but has a number of interesting
features. For a random sequence, at low force and high salt, it will form some
secondary structure, e.g.\ misbonded hairpins. The breaking of this secondary
structure leads to a feature in the force-extension curve (below 10\,pN for
0.5\,M salt).\cite{Huguet10} However, such secondary structure can be
prevented by choosing sequences possessing no complementary bases. If such
sequences have strong stacking, features in the force-extensions curves
associated with force-induced unstacking can be observed.\cite{Chen10b}
However, our model suggests that this transition is not associated with
complete loss of stacking, but rather with a shortening of the length of runs of
stacked bases.\cite{Sulc12} For example, even in the force-bearing
single-stranded sections of the duplex undergoing unpeeling in Fig.\
\ref{fig:stress}(a) one can still see short runs of three or four stacked
bases, where the backbone is able to align itself along the axis of force.

DNA's response to applied twist is dependent on its sign (undertwist is termed
negative supercoiling, and overtwist positive supercoiling), the presence of
tension and sequence.  At low force for long DNAs, the molecule is able to
absorb the twist by writhing to produce plectonemes,\cite{Strick96} such as in
Fig.\ \ref{fig:stress}(c).  For short DNAs or at higher forces, negative
supercoiling can be absorbed in other ways, for example by the formation of a
bubble, as in Fig.\ \ref{fig:stress}(d). One noteworthy feature of such bubbles
in our model is that the two strands twist around each other in the opposite
sense to the DNA double helix. This is so that the bubble absorbs as much of
the negative twist as possible, thus minimizing the number of base pairs that
need to be broken. For sequences possessing an inverted repeat (a stretch of
dsDNA where the sequence reads the same on either strand) the molecule can
absorb the negative supercoiling by forming a double hairpin structure called a
``cruciform'' (Fig.\ \ref{fig:stress}(e)). 
Using oxDNA, we have observed that there is a cooperative but asynchronous
mechanism for the formation of the hairpin arms of the cruciform. The
nucleation of the first hairpin from a bubble that has diffused to the centre
of the inverted repeat makes the nucleation of the second arm much more
likely.\cite{Matek12}

There has also been much interest in the bending fluctuations of 
dsDNA.\cite{Vologodskii13}
Cyclization rates for large DNAs agree well with the predictions of the
wormlike-chain model.\cite{Shore81} However, shorter duplexes (of the order of
the persistence length and less) show enhanced cyclization rates compared to
this model.\cite{Cloutier04,Vafabakhsh12b} It has been suggested that
these enhanced bending fluctuations are due to kinking at sites where transient
bubbles form.\cite{Yan04,Sivak12}

Bending of DNA can also be induced by hybridization of a circular strand with
a linear strand to give a duplex where the two ends of the double helix are
connected by a single strand.\cite{Shroff08,Qu11} As this single-stranded
section becomes shorter, the structure of the double-stranded section changes
from being homogeneously bent to developing a kink due to the formation of a
bubble roughly in the centre of the double-stranded section. For the example
illustrated in Fig.\ \ref{fig:stress}(f), both the bent and the kinked
structures are possible for oxDNA.  For the homogeneously bent case, the
tension in the single-stranded section enhances fraying of the double helix.
However, this tension is significantly reduced on kinking.

The mechanical properties of DNA nanotechnology systems are also beginning to
be probed. For example, single-molecule measurements of the persistence lengths
and torsional moduli of DNA origami multi-helix bundles have recently been
obtained using magnetic\cite{Kauert11} and optical\cite{Pfitzner13} tweezers,
with the extreme rigidities of these origamis making them particularly
suitable for use as linkers in tweezer experiments.\cite{Pfitzner13}

\begin{figure}[t]
\centering
  \includegraphics[width=8.6cm]{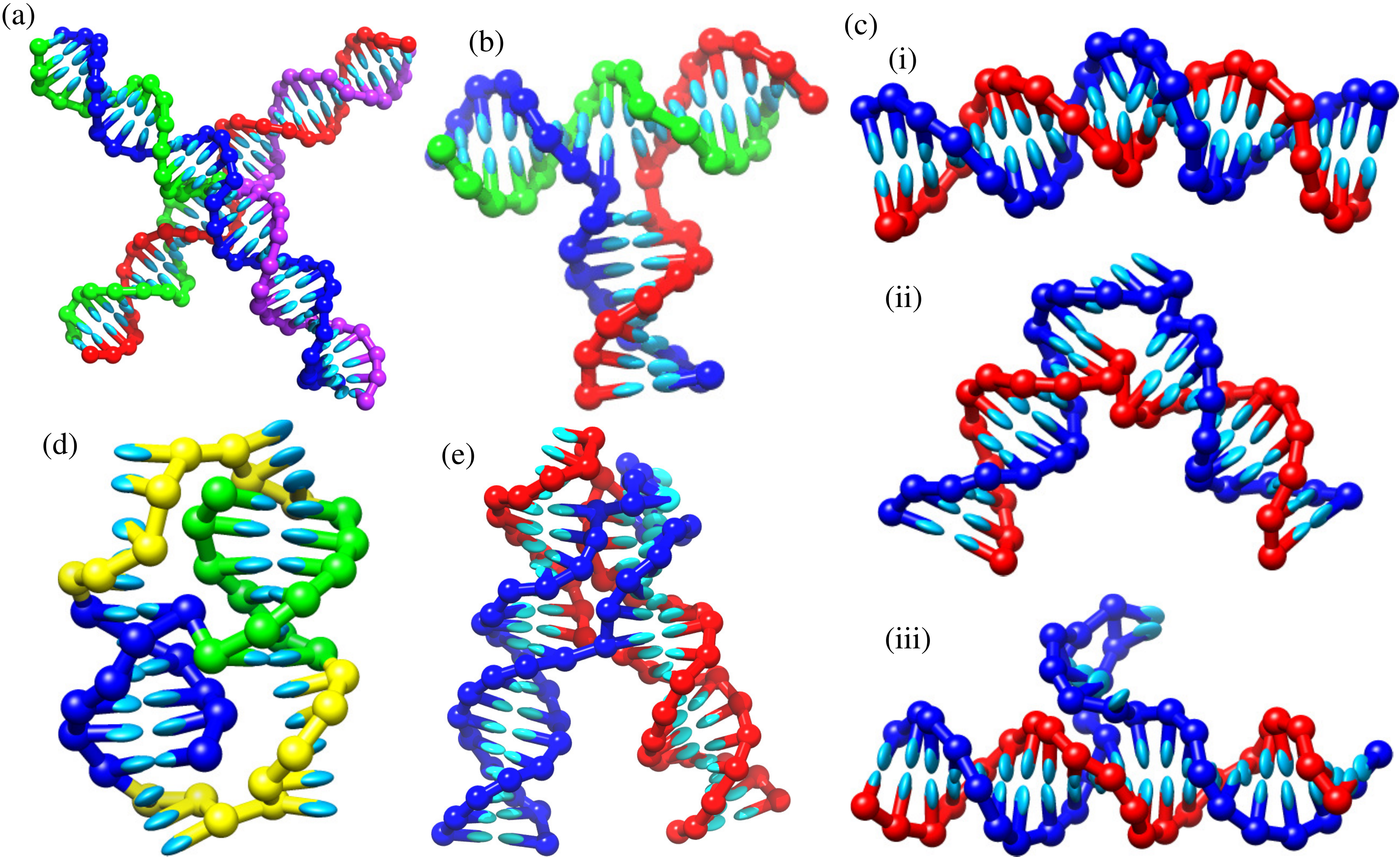}
  \caption{Simple DNA motifs:
(a) a 4-way junction;
(b) a 3-way junction; 
(c) duplexes with a bulge loop of (i) 2, (ii) 5 or (iii) 9 bases;
(d) a pseudoknot and
(e) kissing hairpins. 
}
  \label{fig:structures_intermediate}
\end{figure}

\subsection{Structural DNA Nanotechnology}

As DNA nanotechnological structures basically consist of double helical
sections connected by junctions and single-stranded sections, it will be useful
to consider the properties of such junctions.  The most common junction is a
four-way junction. At the intermediate to high salt typically used in DNA
nanotechnology, the junction does not adopt an open geometry, but instead
adopts a stacked structure, where there is coaxial stacking between the arms at
the junction site, so that there are two (quasi)-continuous helices passing
through the junction. These helices prefer not to be parallel, and adopt an
``X-like'' configuration. Fig.\ \ref{fig:structures_intermediate}(a) shows such
a junction for oxDNA. There are two possibilities for the chiral twist of the
two helices with respect to each other. In our model the junction
preferentially adopts the left-handed form, and although this form has been
seen experimentally,\cite{Nowakowski00} the norm is for the junctions to be
right-handed.\cite{Lilley00} Presumably, oxDNA has insufficient structural
detail to reproduce this local preference. However, this deficiency is likely
to be not too detrimental to the study of the nanotechnological systems, since
the junctions are usually not free to adopt their preferred local structure,
but are constrained by other parts of the nanostructure. In particular, the
helices are often roughly anti-parallel, as, for example, in DNA origami.
However, the internal stresses that result would be expected to be of the
incorrect sign with respect to the twisting of the helices at the junctions.
We are currently exploring these rather subtle effects further in order to 
understand how they affect large DNA nanostructures.

3-way junctions, although somewhat less common, are another basic motif that is
used in DNA nanotechnology,\cite{Seeman82}
for example at the corners of a DNA nanotetrahedron.\cite{Goodman05} 
The two basic proposed types of structure for the junction are a T-shaped
geometry, where two of the arms coaxially stack at the junction, or a geometry
with the arms at approximately 120$^\circ$ and an open arrangement at the
junction without any coaxial stacking.\cite{Lilley00} The competition between
these structures depends on solution conditions and whether there is a
bulge loop at the junction --- this provides the structure with more
flexibility making coaxial stacking easier.  The T-shaped geometry that we find
for oxDNA is illustrated in Fig.\ \ref{fig:structures_intermediate}(b); the
identity of the two arms that are coaxially stacked was found to switch
frequently.

Another common motif that can be used to help control the structure of DNA
assemblies is a bulge loop where one strand has a section of extra
non-complementary bases. Some example structures for duplexes with bulges are
shown in Fig.\ \ref{fig:structures_intermediate}(c) with the structures
reflecting the nature of the stacking in the bulge region. If the bases on
either side of the bulge coaxially stack as in Fig.\
\ref{fig:structures_intermediate}(c)(iii) the duplex is relatively
unperturbed, whereas if the stacking is broken between bases on 
the non-bulge-containing strand the bulge acts as a relatively flexible hinge
allowing larger angular deviations between the two double helical sections
(Fig.\ \ref{fig:structures_intermediate}(c)(ii)). In between these two cases
is Fig.\ \ref{fig:structures_intermediate}(c)(i) where the bases in the bulge
stack with those of the duplex and there are no breaks in the stacking in the
other strand, and a small bend is induced.

\begin{figure*}[t]
\centering
  \includegraphics[width=17cm]{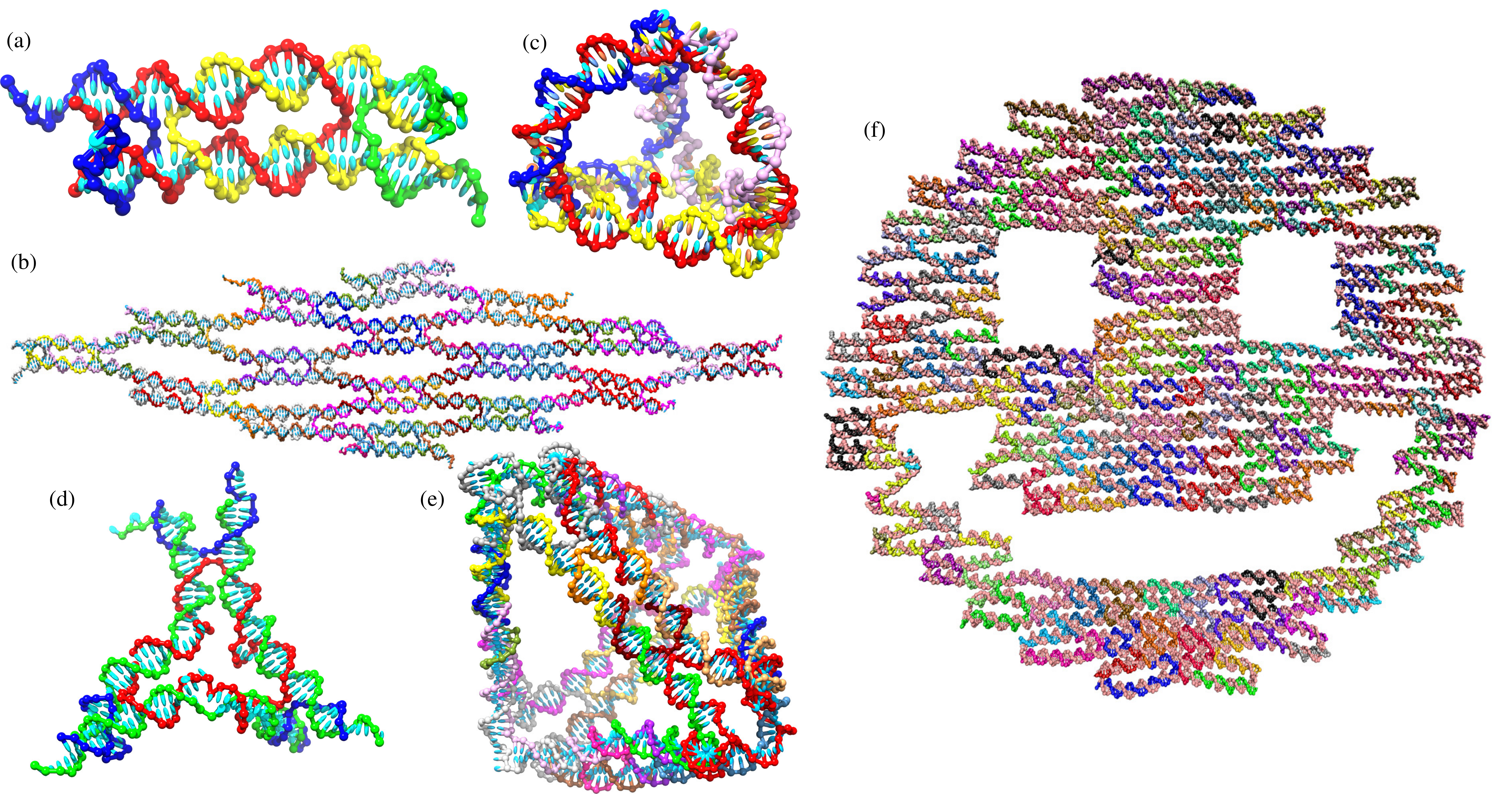}
  \caption{Examples of DNA nanostructures: (a) a double crossover tile, and (b)
a $4\times 4$ array of such tiles adsorbed on a surface, (c) a nanotetrahedron, 
(d) a three-armed star motif and (e) a tetrahedron made up of four such motifs,
and (f) a ``smiley-face'' DNA origami made up of 6196 base pairs
adsorbed on a surface.}
  \label{fig:nanostructures}
\end{figure*}

The connectivity of the double helical sections can also sometimes lead to
pseudoknotted configurations.\cite{Ouldridge11b,Ouldridge13b} The example
pseudoknot illustrated in Fig.\ \ref{fig:structures_intermediate}(d) is
effectively a hairpin with a tail that binds to the loop to form a second
helical section. The thermodynamics of this system are interesting since,
although the two helices in this example are equivalent, the free-energy
profile in Fig.\ \ref{fig:FEL}(b) shows that going from either of the two
possible hairpins to the pseudoknot by forming a second helical section is
thermodynamically less favourable than forming the hairpin. As a consequence,
as the system is cooled down, the system will form a hairpin at 320\,K  but
only at 299\,K does the pseudoknot become most stable.  It is also noteworthy
that secondary structure thermodynamic models find pseudoknots particularly
challenging,\cite{Dirks07} again illustrating the extra insights available
from models with a full three-dimensional representation of the structure.  

One of the first larger-scale self-assembled DNA nanostructures were the 
two-dimensional arrays of double-crossover tiles\cite{Fu93} produced by Winfree
{\it et al.}.\cite{Winfree98} An individual double-crossover tile is
illustrated in Fig.\ \ref{fig:nanostructures}(a), and a small $4\times 4$ array
of such motifs in Fig.\ \ref{fig:nanostructures}(b).  Their self-assembly is
hierarchical with the double-crossover tiles forming at higher temperature, and
then these further self-assembling into arrays 
mediated by interactions between the tiles' short single-stranded ``sticky
ends''. 
The arrays are usually visualized after deposition on a surface, and are then
of course flat. However, their structure and flexibility in solution is less
clear, although the tendency of some tile types to form tubes 
suggests a preference for these sheets to 
curved.\cite{Rothemund04} 
Particularly noticeable in the structure in Fig.\ \ref{fig:nanostructures}(b)
are the gaps that open up between adjacent tiles, and this compares well to the
``rectangular checkerboard'' pattern typically seen in experimental
micrographs, where the gaps are of the order of half of the width of the
tile.\cite{Rothemund04} Our structure confirms that these gaps are the result
of bending of the helices, particularly near to the junctions, but without 
significant loss of stacking at the junctions or strand ends.

We have also begun to explore the assembly of the tiles into arrays by
dynamics simulations. 
As when we considered the hairpin, there will be a temperature window just
below the melting point of the crystals where the
only stable product is the correctly-formed array. However, in this
temperature window there is a ``nucleation'' free-energy barrier to assembly,
and so hysteresis is experimentally observed between assembly and disassembly
of the arrays.\cite{Rothemund04b,Schulman07}  In particular, tiles bound by
a single sticky-end to the rest of the array will not be stable, and have a
significant likelihood of dissociating, before one of its other sticky-ends
either find its partner in the aggregate or binds to a new tile joining the
aggregate.  Consequently, simulating assembly in this regime is challenging,
because many binding events do not lead to growth, and nucleation of the
initial array is especially difficult. In particular, a dimer will not be
stable, and only once a nucleus of four tiles is formed can each tile be bound
by two of its sticky ends. It is far easier to simulate assembly for this
system in a lower temperature range where every sticky-ended association is
stable, and so there is no nucleation free-energy barrier. However, this can
lead to imperfect structures, because, for example, before all the sticky ends
of a tile find their correct binding partner in an aggregate, another tile can
come in and stably bind to one of those partners --- this effect is also
exacerbated by the higher concentrations that one typically uses in
simulations to reduce diffusion times. 
Forward-flux sampling provides a potential means to operate within the
``nucleation'' regime by sampling the successive addition of tiles to a
growing aggregate.

DNA strands can also be designed to form finite structures of a particular
size and shape. Fig.\ \ref{fig:nanostructures}(c) illustrates a
nanotetrahedron designed by the Turberfield group to assemble from four
strands, where each strand runs around one face of the
structure.\cite{Goodman05} In this case, the structure is designed to form
hierarchically, with dimers involving the two edges with no nicks forming
first at higher temperature. These two types of dimers will then themselves
dimerize with each other to form the tetrahedron as the system is cooled
further.

A different route to polyhedra is through the multi-arm motifs developed
particularly in the group of Chengde Mao.\cite{He09,Zhang08,He05,Yan03,He06}
An example three-arm tile is shown in Fig.\ \ref{fig:nanostructures}(d). The
arms have a very similar structure to one half of the double-crossover tiles,
including single-stranded sticky ends that allow the motifs to assemble into
higher-order structures.  To impart flexibility to the motifs, the red strand
in Fig.\ \ref{fig:nanostructures}(d) has three 
bulge loops at the centre of the motif. It is noticeable that for the
configuration in Fig.\ \ref{fig:nanostructures}(d), there is a coaxial
stacking across the bulge in one case leading to a single homogeneously bent
double helix bridging two of the arms, whereas the stacking is broken at the
other two bulges giving rise to a kink. It is also noticeable that the two
helices in an arm are not necessarily parallel, reflecting the preference for
the coaxially stacked helices in a four-way junction to be twisted with
respect to each other.

The interactions and assembly products of these multi-arm motifs can be
controlled in a number of ways, including through the length and number of the
arms, the number of bases in the bulges and the strand concentration.  If the
length of two arms bound to each other is an integer multiple of complete
helical turns, then the two motifs have the same orientation, and so any
non-planarity of the motifs can lead to a build up of curvature, making closed
structures more likely. By contrast if the length is an odd multiple of half
helical turns, the motifs will have opposite orientation, and any non-planarity
will be cancelled out, making planar crystalline arrays more likely. This
approach has been used to create octahedra\cite{He09} and
icosahedra\cite{Zhang08} from 4- and 5-arm motifs respectively, and
honeycomb,\cite{He05} square\cite{Yan03} and hexagonal\cite{He06} lattices from
3-, 4- and 6-arm motifs, respectively.

The case of closed structures with 3-arm motifs is particularly
interesting,\cite{He08} as the structures that result reflect the relative
time scales for internal structural fluctuations in a growing aggregate that
allow closed polygons to form and for addition of new motifs to the growing
structure.  When the bulges are five bases long and the motifs particularly
flexible, tetrahedra form --- it is noticeable from the tetrahedron in Fig.\
\ref{fig:nanostructures}(e) how the unpaired bases in the bulge loops are
stretched out to bridge the helices in the different arms and allow the sharp
back-bending at the vertices of the tetrahedron.  When the bulges are shorter
and the motifs less flexible, larger structures such as dodecahedra and
truncated icosahedra can form, with the larger structure being favoured by
higher concentration, as this increases the rate of addition of new motifs to
a growing structure. We intend to use oxDNA to probe how the length of the
bulge loops affects the rate of polygon closing using forward-flux sampling.

Comparing the two tetrahedra in Fig.\ \ref{fig:nanostructures}(c) and (e), it
is noticeable that the edges of the larger tetrahedron are significantly
straighter; the presence of two linked helices along each edge leads to a
substantial enhancement in their stiffness.

A significant advance in the repertoire of nanostructures that could be
reliably formed was through the development of the DNA origami technique by
Rothemund,\cite{Rothemund06} in which a long single-stranded viral
``scaffold'' DNA is folded up into a structure made of linked parallel double
helices by the addition of lots of short ``staple'' strands.  The iconic
smiley-face origami (Fig.\ \ref{fig:nanostructures}(f)) illustrated that
virtually any arbitrary two-dimensional shape could be formed. The technique
was then further extended to allow three-dimensional\cite{Douglas09} and
curved\cite{Dietz09,Han11} structures to form.

The origami illustrated in Fig.\ \ref{fig:nanostructures}(f) shows that our
model is able to capture one of the most basic structural features of DNA
origami, namely the ``weave'' pattern where the spacing between the helices
opens up in between the junctions. 
However, the structure of origamis are virtually always probed when bound to a
surface, and much less is known experimentally about their structure in
solution, although the CanDo package can provide a useful
guide.\cite{Castro11,Kim12}  We find that two-dimensional origamis can have
quite large structural fluctuations, and need not be planar on average.  As
well as the weave pattern, we find that the origamis can be ``corrugated'' in
the plane perpendicular to the origami, as has also been seen in the recent
cryo-EM structure of a three-dimensional origami.\cite{Bai12} The original
origami of Rothemund are almost certainly twisted in solution, because the
design requires the helical repeat length to be 10 2/3
base pairs, and so the mismatch between that and DNA's natural helical repeat
of about 10.5 bases per turn leads to twist, as has been illustrated for long
three-dimensional origami ribbons.\cite{Dietz09}
For oxDNA, we find that the overall twist also reflects the effects of nicks,
and junctions on the local twist angle at the corresponding base-pair steps. 

We note that a programme to take output from cadnano,\cite{Douglas09b} a
commonly used DNA origami design tool, and turn it into an input file for oxDNA
is available at the oxDNA website.\cite{oxDNA}

Experiments have shown that there is hysteresis between the self-assembly and
melting of DNA origamis, that the self-assembly can occur over a relatively
narrow temperature window, and that if the system is quenched below this
assembly window the quality of the resulting origami
decreases.\cite{Sobczak12,Yang12} These results suggest that near the melting
temperature there is a nucleation free-energy barrier to origami formation;
this barrier probably reflects the decrease in conformational entropy of the
remaining single-stranded sections of the scaffold strand as staple strands
bind making the binding of each successive staple more thermodynamically
favourable.  We have begun to use oxDNA to explore the self-assembly of small
DNA origamis, but, similar to the assembly of double-crossover tiles, direct
simulations of assembly have to be in a temperature regime where the assembly
process is downhill in order to occur in a reasonable amount of computer time.
However, we then see kinetic trapping in configurations where two copies of
the same staple have both bound to the scaffold, thus blocking either staple
strand fully bonding to the scaffold. In order to avoid such problems, one
would need to work at a temperature where the binding of the first domain of
each staple is unfavourable, so that singly-bound blocking strands would melt
away, but this regime would require rare-event techniques, such as
forward-flux sampling to be used.

\subsection{Active DNA nanotechnology}

In DNA nanodevices, activity is often achieved through what are termed
strand-exchange or displacement reactions, in which an ``incumbent'' strand
that is partially complementary to a ``substrate'' strand is replaced by an
``invading'' strand that is able to form more base pairs with the substrate.
Figure \ref{fig:basic_dynamics}(b) illustrates this process. Interestingly, the
rate of displacement is found to initially increase exponentially with the
length of the toehold (the number of unpaired bases in the substrate-incumbent
complex) before plateauing at 5 or 6 bases.\cite{Zhang09b} In contrast to
simple kinetics schemes based on nearest-neighbour thermodynamic models, oxDNA
is able to quantitatively reproduce the $10^{6.5}$-fold acceleration of the
rates with increasing toehold length,
because it can capture the free-energetic effects of the overcrowding at the
junction between the invading and incumbent strands 
and the dynamical consequences of the greater structural rearrangement
necessary for displacement to progress compared to toehold
melting.\cite{Srinivas13}

\begin{figure}[t]
\centering
  \includegraphics[width=8.6cm]{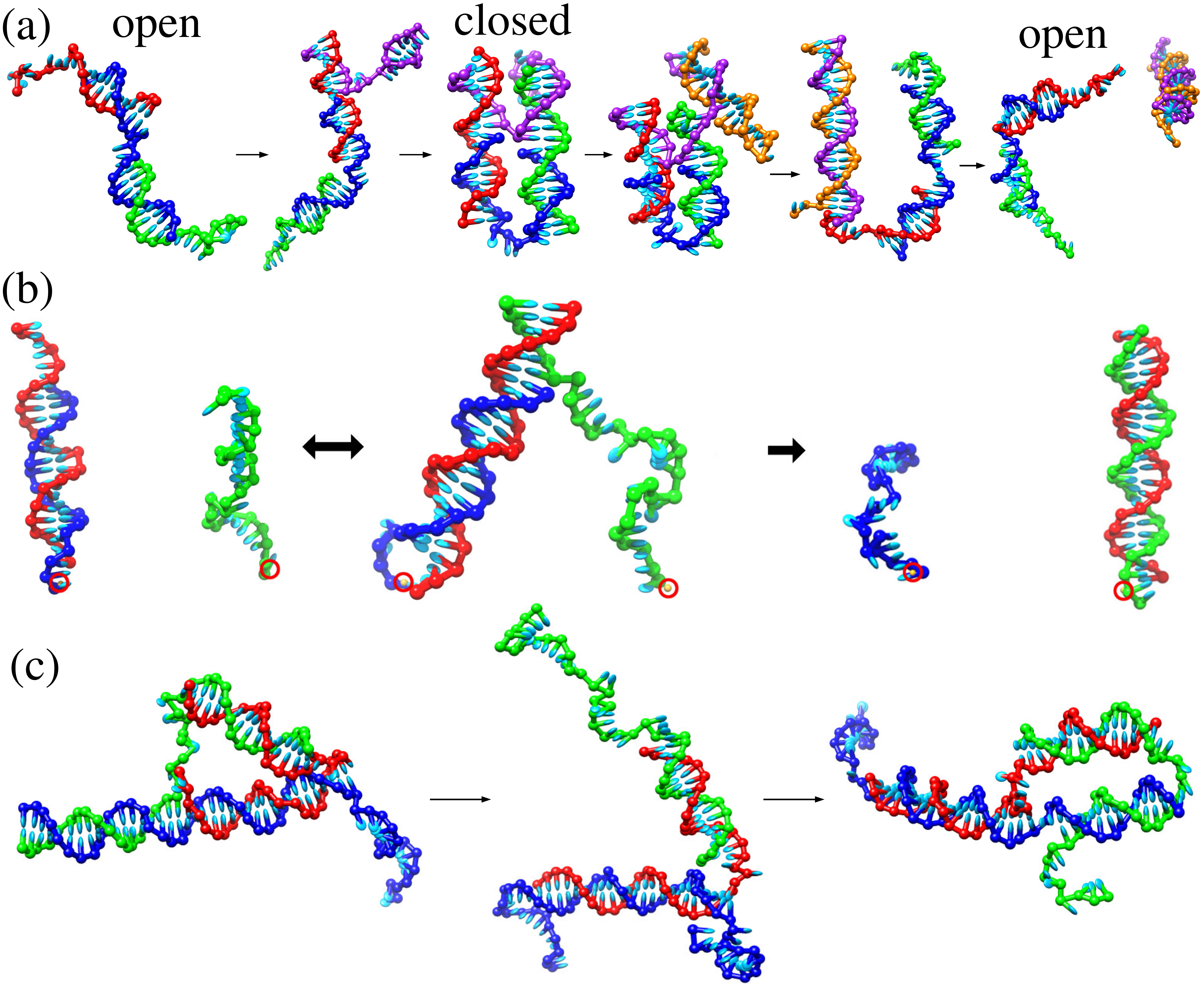}
  \caption{Examples of DNA nanodevices: (a) nanotweezers transforming between
their open and closed states in response to the addition of fuel (purple) and 
anti-fuel (gold) strands, (b) a burnt-bridges walker moving between two 
stators, and (c) a two-footed DNA walker taking a step along a single-stranded 
track (blue). In (b) the red circles indicate the points of attachment to the
substrate.
}
  \label{fig:nanodevices}
\end{figure}

One of the first nanodevices to use displacement was the DNA nanotweezers
developed by Yurke {\it et al.},\cite{Yurke00} an example of which is
illustrated in Fig.\ \ref{fig:nanodevices}(a).\cite{Ouldridge10} The
nanotweezers cycle between ``open'' and ``closed'' configurations through the
``clocked'' addition of complementary single strands. Using oxDNA, we have 
been able to characterize the free-energy landscape associated with 
a complete cycle of the nanotweezers (Fig.\ \ref{fig:FEL}(d)).
As the two strands would preferentially hybridize with each other rather than
open or close the tweezers, their addition is necessarily sequential in time.
One way round this restriction is to prepare these ``fuel'' strands in a form
that prevents them from hybridizing directly, but where the nanodevice can extract
work by catalysing their hybridization. This can be achieved if the strands
are designed to form hairpins.\cite{Bois05,Green06} Although the hairpin loops
can start to hybridize, the topology prevents full hybridization (Fig.\
\ref{fig:FEL}(c))---the linking number of the two loops must be conserved so
for every time the two loops wrap around each other to form the B-DNA double
helix, they must also wrap around each other in the opposite sense.  The
structure of the ``kissing complex'' formed by two such hairpins is
illustrated in Fig.\ \ref{fig:structures_intermediate}(e).  Perhaps contrary
to expectation, oxDNA suggests that it is favourable for the loops to form two
double-helical sections adjacent to each hairpin stem rather than a single
hybridized region.\cite{Romano12b}

An alternative approach to introduce dynamical activity is to use a modified
restriction enzyme to nick one of the strands in a double helical section, such
that it is favourable for at least one of the two fragments to
dehybridize. This approach has been used to create a ``burnt-bridges'' motor
that walks along a track of single-stranded stators attached to some substrate
(be it a single double helix\cite{Bath05} or a DNA origami\cite{Wickham11}) and
is illustrated in Fig.\ \ref{fig:nanodevices}(b).  The enzyme nicks the
stator-cargo complex to reveal a toehold that allows the cargo to be
transferred to the next stator through a displacement reaction.  Thus, the
track is modified as the motor progresses, hence the ``burnt-bridges'' name. We
have used our model to probe the free-energy landscapes for the displacement
process, and particularly the effect of stator separation.\cite{Sulc13} As the
stators become increasingly separated, there is an increasing free-energy
barrier to displacement because of the tension that builds up in the
intermediate as it is stretched between the binding sites of the two stators
(Fig. \ref{fig:FEL}(e)).  This effect is potentially useful for tracks where
one wants the cargo to select one of two paths,\cite{Wickham12} because as the
probability of displacement being successful diminishes (due to the large
free-energy barrier), the relative rates of taking either path become more
sensitive to the toehold binding strengths.

We have also modelled a prototype two-footed walker developed in the
Turberfield group that walks along a single-stranded track. There are two
versions of the walker, one that makes use of nicking enzymes\cite{Bath09} and
one that uses hairpins as fuel,\cite{Green08} the former being illustrated in
Fig.\ \ref{fig:nanodevices}(c). The motor is designed so that the two-feet
overlap, and hence expose a toehold that allows either of the feet to be
selectively detached from the track by displacement using an appropriate fuel
strand. Therefore, in the presence of the fuel that lifts the back foot, the
walker should walk forwards. Our modelling suggests that there is also a
potential bias, stemming from the geometry of the walker, for the lifted foot
to reattach in the forward direction, especially when moderate tension is
applied to the track.\cite{Ouldridge13} However, we also found that the walker
had a tendency to ``overstep'', and attach to a binding site not adjacent to
the stationary foot, which is a potential problem because the feet then no
longer overlap, making it harder for the feet to be lifted by the fuel strands
and eliminating the intended bias for lifting the back foot. Applying moderate
tension to the track was found to make recover from the overstepped state
easier.  This study probably provides the first example of optimization
strategies for a DNA nanodevice being suggested based on
simulations.\cite{Ouldridge13}

\section{Conclusions}

Molecular simulation has potentially much to offer the fields of DNA
nanotechnology and DNA biophysics. However, for this promise to be realized,
accurate and robust coarse-grained models of DNA are required in order to
address the potentially long time and length scales involved.  After a
relatively late start compared to coarse-grained simulations of biomolecules
such as proteins and lipids, coarse-grained modelling of DNA has received
considerable interest over the last few years, with many possible models now
available in the literature and an increasing number of applications using
these models.  However, for quite a number of these models, their fundamental
behaviour has not been sufficiently tested to be confident of their ability to
capture a wide range of phenomena. Furthermore, many models involve choices
concerning the form of their interactions that 
limit their ability to exhibit the physical properties that are most relevant
to applications to the self-assembly processes associated with DNA
nanotechnology, namely a realistic description of the structure, thermodynamics
and mechanics of both dsDNA and ssDNA. 

By contrast, the oxDNA model, which was developed by the authors and is the
focus of the second-half of this perspective article, was specifically
designed to capture the biophysical processes that are essential to
self-assembling DNA nanotechnology.  As well as accurately reproducing the
structural, thermodynamic and mechanical properties that were involved in the
fitting process, oxDNA can be simulated on the diffusive time scales relevant
to self-assembly processes, even for systems containing thousands of
nucleotides.  Furthermore, the power of the model is exemplified by the
quantitative reproduction of phenomena to which the model was not fitted, most
notably the kinetics of toehold-mediated strand-exchange and the
overstretching force of duplex DNA. Its utility is further illustrated by the
wide range of examples to which the model has been applied, as outlined in
Section \ref{sect:results}. It is able to provide significant physical insight
into fundamental dynamic processes involving DNA, such as hybridization,
strand exchange and hairpin formation, and the response of DNA to mechanical
stress, be it stretching, twisting or bending, as well as providing an
excellent description of the structural properties of DNA nanosystems.
Moreover, its ability to capture non-trivial geometric features of nanodevices
that are not accessible to secondary-structure thermodynamic models gives it a
potentially important role in guiding the design of DNA nanodevices, as has
been illustrated for a prototype two-footed DNA walker.  OxDNA has also been
used to probe the liquid-crystalline phase behaviour of concentrated solutions
of short DNA duplexes.\cite{deMichele12}

Having illustrated oxDNA's achievements, it is also important to be open about
its deficiencies and limitations. Probably the main limitation is that the
model is fitted to one relatively high salt concentration (0.5\,M), albeit one
that is typical of the high ionic strengths usually used in DNA
nanotechnology.  Furthermore, the sequence dependence in the model is limited
to just the thermodynamic properties. Thus, there is no sequence-dependent
elastic behaviour, which is probably less important for nanotechnology, but
may be relatively more important for applications in biology.  Additionally,
the only form of base pairing allowed is Watson-Crick, and so alternative
forms of DNA, such as G-quadruplexes and triple-stranded DNA, cannot be
modelled. The symmetric nature of the oxDNA helix is a further simplification,
but one that we are currently addressing due to its potential structural
effects on DNA nanostructures. 
Finally, the property most relevant to DNA nanotechnology that oxDNA is not
able to reproduce is probably the structure of four-way junctions for which
oxDNA has a preference for the isomer with the opposite chirality to that
usually observed experimentally.

In summary, coarse-grained modelling of DNA has now reached an exciting stage.
Models are now available that allow DNA nanosystems to be systematically and
accurately probed, thus opening up the field of DNA nanotechnology to
molecular simulations. Even complete DNA origamis with over ten thousand
nucleotides can be structurally characterized. Probing the self-assembly
mechanisms of such large nanostructures is still a real challenge for
simulations because of the long time scales involved, but it is envisaged that
through the combination of GPU computing and advanced rare-event techniques
simulations of the self-assembly of structures with of the order of a thousand
nucleotides will soon be possible.

\section*{Acknowledgements} 
The authors are grateful to the Engineering and
Physical Sciences Research Council, University College (Oxford) and
the German Academic Exchange Service (DAAD)
for financial support.
PS is grateful for the award of a Scatcherd European Scholarship
and RMH for a National Science Foundation Graduate Research Fellowship.

\footnotesize{
\providecommand*{\mcitethebibliography}{\thebibliography}
\csname @ifundefined\endcsname{endmcitethebibliography}
{\let\endmcitethebibliography\endthebibliography}{}

}

\end{document}